\documentclass[twocolumn,aps,superscriptaddress]{revtex4}
\usepackage{graphicx}
\usepackage{float}
\usepackage{dcolumn}
\usepackage{bm}
\usepackage{color}
\usepackage{amsmath}
\usepackage{amssymb}
\usepackage{amsfonts}
\usepackage{esint}
\usepackage{times}
\usepackage{xcolor}
\usepackage{phaistos}
\usepackage{braket}
\usepackage{comment}

\usepackage{pdfpages}

\usepackage[colorlinks,linkcolor=blue,anchorcolor=blue,urlcolor=blue,citecolor=blue]{hyperref}

\begin{document}

\title{Mitigation of quantum crosstalk in cross-resonance based qubit architectures}

\author{Peng Zhao}
\affiliation{Tongling, Anhui 244000, China}
\affiliation{Beijing Academy of Quantum Information Sciences, Beijing 100193, China}

\date{\today}

\begin{abstract}

The Cross-resonance (CR) gate architecture that exploits fixed-frequency transmon qubits and
fixed couplings is a leading candidate for quantum computing. Nonetheless,
without the tunability of qubit parameters such as qubit frequencies and couplings, gate operations can be limited by the
presence of quantum crosstalk arising from the always-on couplings. When increasing system sizes, this can
become even more serious considering frequency collisions caused by fabrication uncertainties.
Here, we introduce a CR gate-based transmon architecture with passive mitigation of both quantum crosstalk and
frequency collisions. Assuming typical parameters, we show that $ZZ$ crosstalk can be suppressed
 while maintaining $XY$ couplings to support fast, high-fidelity CR gates.
The architecture also allows one to go beyond the existing literature by extending the operating regions
in which fast, high-fidelity CR gates are possible, thus alleviating the frequency-collision issue.
To examine the practicality, we analyze the CR gate performance
in multiqubit lattices and provide an intuitive model for identifying and mitigating the dominant
source of error. For the state-of-the-art precision in setting frequencies, we further
investigate its impact on the gates. We find that $ZZ$ crosstalk and frequency collisions can be largely
mitigated for neighboring qubits, while interactions beyond near-neighbor qubits can introduce new
frequency collisions. As the strength is typically at the sub-MHz level, adding weak
off-resonant drives to selectively shift qubits can mitigate the collisions. This work could be
useful for suppressing quantum crosstalk and improving gate fidelities in large-scale quantum processors
based on fixed-frequency qubits and fixed couplings.

\end{abstract}

\maketitle


\section{Introduction}\label{SecI}

Achieving high-fidelity qubit operations in scaling up qubit architectures is one of the most
fundamental challenges in quantum computing \cite{Martinis2015}. With demonstrations of long
coherence times, ease of control, and flexibility in design, superconducting qubits hold great
promise for implementing large-scale quantum processors. Accordingly, diverse qubit architectures
based on superconducting qubits have been developed during the past two
decades \cite{Krantz2019,Blais2021}. However, until now, very few of those have demonstrated
significant improvements in both system sizes and gate fidelities. Besides qubit decoherence, the main
obstacle is due to the presence of quantum crosstalk arising from residual qubit-qubit couplings \cite{Bialczak2011}. This highlights
the importance of understanding the quantum crosstalk and figuring out how to address this issue
for building large-scale superconducting quantum processors.

The cross-resonance (CR) gate is an all-microwave-controlled two-qubit gate for superconducting qubits, which is
implemented by driving the control qubit at the frequency of the target qubit \cite{Paraoanu2006,Rigetti2010,de Groot2010,Chow2011}, and
has been shown to have considerable potential for building large-scale quantum processors based on transmon qubits \cite{Koch2007}.
For instance, the past decade has witnessed tremendous progress toward improving both the system size
and gate performance for the CR-based transmon architecture, which has led to the demonstrations of
small-scale error-correction codes \cite{Chen2022,Sundaresan2023} and the exploration of quantum
advantage before fault tolerance \cite{Kim2023}. To be more specific, the system size can
reach the 100-qubit level \cite{Kim2023}, and the typical single-qubit gate error is close to $10^{-4}$ while
the two-qubit gate error is below $10^{-2}$ \cite{Kim2023,Chen2022,Sundaresan2023,Jurcevic2021}. 

\begin{figure}[tbp]
\begin{center}
\includegraphics[keepaspectratio=true,width=\columnwidth]{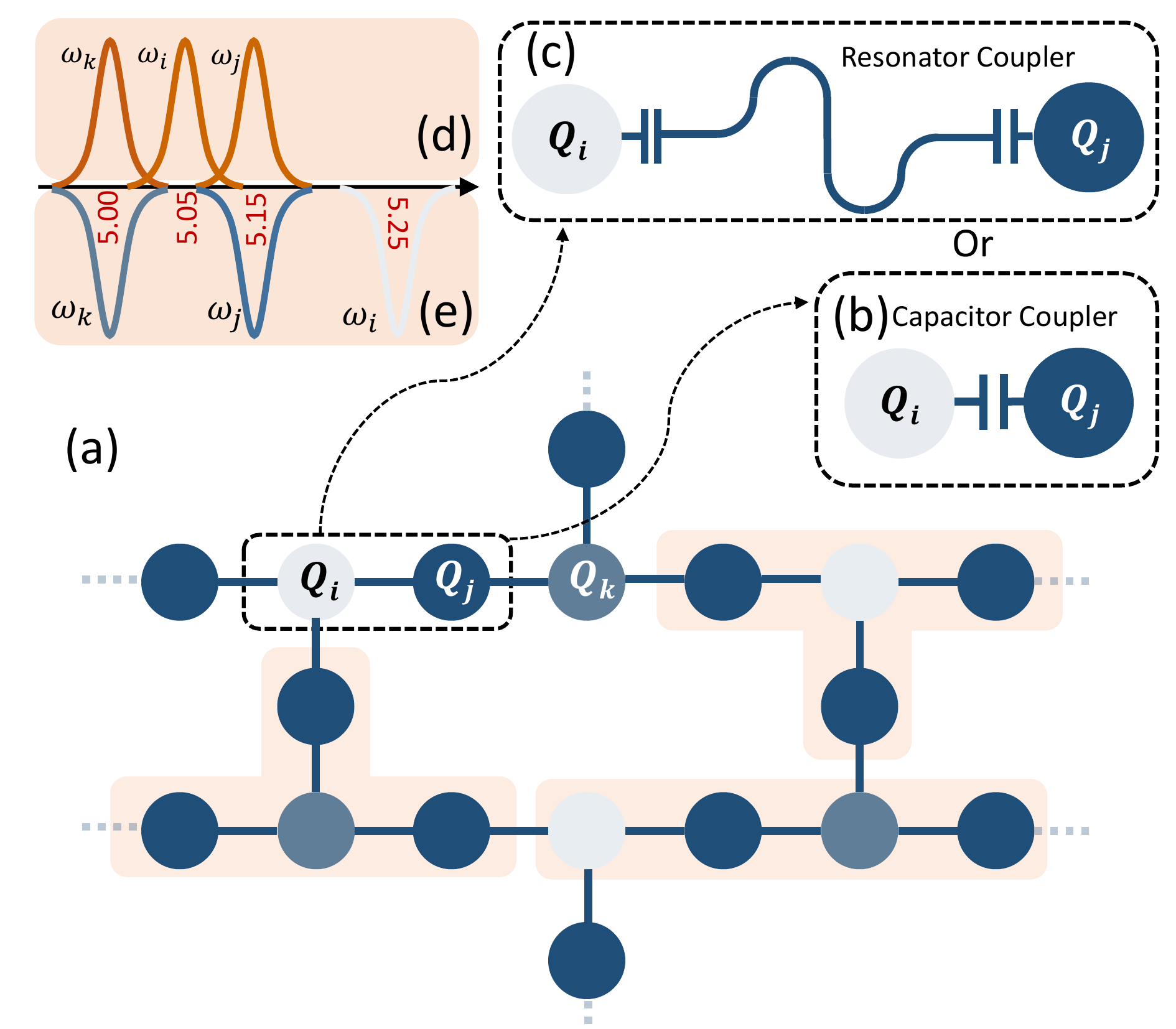}
\end{center}
\caption{(a) The heavy-hexagonal qubit lattice. The circles denote transmons while the edges
represent couplers, which could be a capacitor (b) or a resonator (c).
Here, a three-frequency pattern indicated in three colors (dark blue, light grey, and dark grey) is
employed to avoid frequency collisions. For CR architectures, the dark blue circles denote the
control qubits, e.g., $Q_{j}$ with the typical frequency $\omega_{j}$, and the light (dark) grey
circles denote the target qubits, e.g., $Q_{i(k)}$ with the typical frequency $\omega_{i(k)}$.
(d) Typical distributions of qubit frequencies in the existing literature, where to ensure fast
gates, the control qubit frequency is higher than the target qubit
frequency ($\omega_{j}>\omega_{i}>\omega_{k}$), giving positive control-target detunings. (e) Typical
frequency allocation for the proposed architecture, where qubits are coupled via a
resonator, as in (b), and the three distinct frequencies satisfy the
condition $\omega_{i}>\omega_{j}>\omega_{k}$, allowing for further mitigating
frequency collisions. This is enabled by an appealing feature of the proposed architecture, e.g., even
under conditions of negative control-target detunings, high-fidelity CR gates
can still be achieved with gate speeds comparable to those with positive detunings.}
\label{fig1}
\end{figure}

The progresses mainly benefit from the appealing feature of the CR architecture, i.e., universal qubit operations are
solely based on microwave drives \cite{Chow2012}. Compared to qubit architectures with tunable
elements \cite{Krantz2019,Blais2021}, this feature supports the practical advantage offered by combining the long
coherence time of non-tunable elements such as single-junction transmons \cite{Place2021,Wang2022,Gordon2022} and the low
overhead of all-microwave control \cite{Krantz2019,Chow2012,Sheldon2016}. This eventually yields the most widely
explored CR architecture wherein fixed-frequency transmons are fixedly coupled via a coupling
capacitor \cite{Patterson2019,Paik2020} or a bus resonator \cite{Chow2012,Chow2015}, as shown in Fig.~\ref{fig1}. However, as
one might expect, there is no such thing as a free lunch: the lack of frequency-tunable elements also means that, unlike qubit
architectures with tunable elements, generally, there is no active approach available for mitigating undesirable effects, such
as qubit frequency collisions owing to the inaccuracy in setting qubit frequencies \cite{Gambetta2017,Brink2018,Hertzberg2021,Kreikebaum2020} and
quantum crosstalk arising from residual couplings and the weak anharmonicity of transmons \cite{Sheldon2016,DiCarlo2009,Gambetta2012a,McKay2019,Malekakhlagh2020,Cai2021,Heya2023}. To
this end, within the CR architecture, one has to devise passive approaches to addressing these challenges.

For the CR architecture, quantum crosstalk can be manifested in many forms, such as $ZZ$
crosstalk \cite{Sheldon2016,DiCarlo2009,Gambetta2012a,McKay2019}, next-neighboring interactions \cite{Malekakhlagh2020,Cai2021,Heya2023}, and undesirable multiphoton transitions \cite{Malekakhlagh2020,Heya2023}. More importantly, it can become even more serious when qubit frequency collisions
happen \cite{Malekakhlagh2020,Cai2021,Heya2023}. To alleviate such concerns, considerable efforts have been made, such as improving the
accuracy in setting qubit frequencies \cite{Hertzberg2021,Kreikebaum2020}, optimizing
frequency allocation \cite{Brink2018,Hertzberg2021,Kreikebaum2020,Morvan2022}, and decreasing
qubit connectivity \cite{Hertzberg2021,Chamberland2020}, as shown, for example, in Fig.~\ref{fig1}. Basically, as the quantum crosstalk comes from
the always-on interqubit coupling, lowering the strength can suppress it. However, this will lead to
slow two-qubit gates, increasing gate errors from qubit decoherence. Thus, it is highly desirable to find a
better solution to address this crosstalk issue without sacrificing gate speeds and fidelities. Going
beyond the CR architecture with fixed-frequency transmons and fixed couplings, several
architectures have been proposed for addressing this issue, such as coupling qubits with
opposite-sign anharmonicity \cite{Ku2020,Zhao2020,Xu2021} or introducing frequency-tunable
elements \cite{Chavez-Garcia2022,Xu2023}. However, generally, these
approaches will introduce new decoherence channels and increase the control complexity. Within the present
CR transmon architecture, several alternative approaches have also been developed, such as using
optimized pulses \cite{Jurcevic2021,Sundaresan2020}, the ac-Stark shift \cite{Xu2021,Mitchell2021,Wei2022}, and
the multi-path coupler \cite{Kandala2021,Zhao2021}, yet it is still an outstanding challenge for suppressing quantum crosstalk while
maintaining fast-speed, high-fidelity gates in a robust and scalable manner.

In this work, we introduce a CR architecture based on fixed-frequency transmons and fixed couplings, which
can passively mitigate both quantum crosstalk and frequency-collision issues. In such architecture, transmon
qubits are coupled by a bus resonator, as shown in Fig.~\ref{fig1}(c). In particular, unlike the setting in
other existing literature \cite{Chow2012,Sheldon2016,Chow2015,Goerz2017}, here the qubit-resonator detuning is comparable
to the qubit anharmonicity $\alpha_{q}$ and the qubit-resonator coupling $g_{rq}$ is more than about an
order of magnitude smaller than the anharmonicity, i.e., $|\alpha_{q}/g_{rq}|\gtrsim 10$, retaining the
dispersive condition. For the typical transmon anharmonicity of $-330\,{\rm MHz}$, here, the qubit-resonator
coupling is below $30\,\rm MHz$. It is, in this sense, that we nickname it the \emph{'lightweight'} resonator coupler.

Within the architecture, the \emph{lightweight} coupler allows one to address two main challenges
in scaling up qubits, i.e., the speed-fidelity tradeoff imposed by $ZZ$ crosstalk and frequency
collisions due to fabrication uncertainty. For the former, as the coupler-mediated $XY$ and $ZZ$
couplings are mainly enabled by virtual transitions in subspaces with different excitations, i.e., one-excitation
and two-excitation subspaces, respectively, this enables one to engineer quantum interference to mitigate $ZZ$ coupling
while retaining $XY$ coupling to support fast-speed CR gates \cite{Kandala2021,Zhao2021,Goerz2017,Jin2021,Li2022,Zhao2022b}.
For the latter, to ensure fast CR gates, the other existing literature often favors the operating regime with positive
control-target detunings \cite{Malekakhlagh2020,Tripathi2019,Ware2019,Magesan2020}. By contrast, here, even under conditions
of negative control-target detunings, high-fidelity CR gates can still be achieved with gate speeds comparable to
those with positive detunings. As a result, besides the positive detuning regions, the proposed architectures allow one to go
beyond the existing literature by operating the qubit system in negative detuning regions, thereby mitigating the
frequency-collision issue further, as illustrated, for example, in Figs.~\ref{fig1}(d) and~\ref{fig1}(e).

In addition, to assess the practicality, we systematically analyze the performance
of the direct CX ($\rm CNOT$) gate \cite{Jurcevic2021,Malekakhlagh2022} on multi-qubit lattices and study the effect of
the qubit frequency uncertainty on the gates, assuming state-of-the-art precision
in setting frequencies \cite{Hertzberg2021}. We find that while $ZZ$ crosstalk and frequency collisions for
neighboring qubits can be largely mitigated, frequency collisions beyond nearest-neighbor qubits
can exist due to the presence of sub-MHz couplings, and can degrade gate
performance. In particular, the frequency collisions could be classified into two major
types: Type-1 is the \emph{static} frequency collision, such as on-resonance couplings of next-neighboring
qubits can lead to the strong state hybridization, degrading individual qubit addressability; Type-2 is
the \emph{dynamic} frequency collision, e.g., CR drive-induced multiphoton
transitions. To address such collisions, we provide an intuitive model mainly based on the
dressed state picture for identifying the primary error source. Then, we show that by optimizing system parameters
or adding weak off-resonant drives to selectively shift the qubits \cite{Mitchell2021,Wei2022,Noguchi2020,Xiong2022,Ni2022,Zhao2022c,Wang2023}, the
leading frequency collisions can be avoided.

\begin{table*}[!tbp]
\caption{Typical qubit parameters and approximations used in analyzing CR gate-based transmon architectures.
Here, the qubit parameters are similar to that extracted from experimental works and the coupling strength is given
with the reference frequency of $5\,\rm GHz$. (a) Qubits are coupled directly via a capacitor \cite{Paik2020}.
(b) Qubits are coupled via a resonator \cite{Sheldon2016}. Here, to account for results in Ref.~\cite{Sheldon2016}, a direct
coupling due to an effective capacitance is also taken into consideration \cite{Ku2020,Galiautdinov2012}.
(c) Qubits are coupled via a multipath coupler \cite{Kandala2021,Zhao2021,Yan2018}. (d) Qubits are coupled by a \emph{lightweight} resonator coupler
introduced in the present work. Additionally, to reduce computation costs while maintaining sufficient accuracy, especially for
analyzing multi-qubit systems, the Rotating-wave approximation (RWA) is employed for (b) and (d).\label{tab1:CR architectures}}
\begin{ruledtabular}
\begin{tabular}{c c c c c c}
 Layout & Capacitor coupler$^{(a)}$ &  Resonator coupler$^{(b)}$ & Multipath coupler$^{(c)}$ & \emph{Lightweight} resonator coupler$^{(d)}$ &   \\
\hline
qubit-resonator coupling, $g_{rq}/2\pi\,{\rm (MHz)}$        &              $0$  &             $78$       &             $78$      &             $25$       &  \\

qubit-resonator detuning, $\Delta_{q}/2\pi\,{\rm (GHz)}$              &              $-$  &         $1.0$       &         $1.0$      &       $0.25$    &  \\

qubit-qubit coupling, $g_{12}/2\pi\,{\rm (MHz)}$        &          $\sim 2$   &          $2g_{r1}g_{r2}/\omega_{r}$      &      6      &   $0$        &       \\

qubit anharmonicity, $\alpha_{q}/2\pi\,{\rm (MHz)}$            &          $-330$   &            $-330$      &        $-330$      &           $-330$     &     \\

RWA           &          with   &            without      &        without      &          with     &     \\

\end{tabular}
\end{ruledtabular}
\end{table*}

The rest of the paper is organized as follows. In Sec.~\ref{SecIIA}, we first provide a brief overview
of CR-based transmon architectures and then discuss $ZZ$ crosstalk and frequency collisions.
In Sec.~\ref{SecIIB}, we give detailed descriptions of the introduced
architecture, mainly focusing on $ZZ$ suppression and revisiting the negative control-target detuning
regime, which is previously identified as a slow gate regime. Within the architecture and assuming
typical parameters, in Sec.~\ref{SecIII}, we numerically analyze the performance of
CX gates, including both isolated and simultaneous gates, on multiqubit lattices. Moreover, we also examine
the impact of frequency uncertainty on gates. In Sec.~\ref{SecIV}, we provide discussions of frequency
collisions and the resulting challenges to be faced when scaling up the architecture.
Finally, in Sec.~\ref{SecV}, we provide a summary of our work.

\section{CR architecture with \emph{lightweight} resonator coupler }\label{SecII}

In this section, for easy reference and to set the notation, we first briefly review the basic principle of
CR gates and discuss the issues of ZZ crosstalk and the frequency collision in CR gate-based transmon architectures. Then, we
give a detailed description of the introduced architecture and show how it can address the two issues.

\subsection{Overview of the CR gate-based transmon architecture}\label{SecIIA}

\subsubsection*{Fixed-coupled transmons and effective Hamiltonians}\label{SecIIA1}

The CR gate is an all-microwave-activated two-qubit gate for fixed-coupled qubits, and within the
CR architecture based on fixed-frequency transmons, qubits can be fixedly coupled via a
coupling capacitor or a bus resonator, as depicted in Figs.~\ref{fig1}(b) and~\ref{fig1}(c). Note here that
for real qubit devices with capacitive coupling, besides the desired couplings,
stray couplings due to, such as stray capacitances or effective capacitances mediated by
coupling circuits \cite{Ku2020,Galiautdinov2012,Yan2018,Yanay2022}, can exist and have non-negligible
contributions to interqubit couplings. By considering this, Table~\ref{tab1:CR architectures} lists
four main types of architectures for coupling transmons, which can be described by the
static Hamiltonian (hereinafter, $\hbar=1$)
\begin{equation}
\begin{aligned}\label{eq1}
H=H_{0}+H_{int},
\end{aligned}
\end{equation}
\begin{equation}
\begin{aligned}\label{eq2}
H_{0}=\sum_{q=1,2}\big(\omega_{q}a_{q}^{\dagger}a_{q}+\frac{\alpha_{q}}{2}a_{q}^{\dagger}a_{q}^{\dagger}a_{q}a_{q}\big)+\omega_{r}a_{r}^{\dagger}a_{r}.
\end{aligned}
\end{equation}
Here $\omega_{q}$ and $\alpha_{q}$ are the frequency and the anharmonicity of transmon
$Q_{q}$ with the annihilation (creation) operator $a_{q}\,(a_{q}^{\dagger})$, and $\omega_{r}$ is the
frequency of the resonator with the annihilation (creation) operator $a_{r}\,(a_{r}^{\dagger})$. The
interaction Hamiltonian is
\begin{equation}
\begin{aligned}\label{eq3}
H_{int}=\sum_{\substack{q,p=1,2,r\\q\neq p}}g_{qp}(a_{q}+a_{q}^{\dagger})(a_{p}+a_{p}^{\dagger}),
\end{aligned}
\end{equation}
where $g_{r1}$ ($g_{r2}$) denotes the strength of the coupling between the resonator (R) and qubit $Q_{1}$ ($Q_{2}$),
while $g_{12}$ is the direct qubit-qubit coupling strength.

Truncated to the computational subspace, the static system Hamiltonian can be approximated by an effective
Hamiltonian based on the two-level qubit model and is given as \cite{Zhao2021,Magesan2020}
\begin{equation}
\begin{aligned}\label{eq4}
\widetilde{H}=\tilde{\omega}_{1}\frac{ZI}{2}+\tilde{\omega}_{2}\frac{IZ}{2}+J_{xy}\frac{XX+YY}{2}+\zeta_{zz}\frac{ZZ}{4},
\end{aligned}
\end{equation}
where $(X,Y,Z,I)$ denote the Pauli operators and $\tilde{\omega}_{q}$ is
the dressed qubit frequency. Here, the third term describes the $XY$ coupling resulting from the direct qubit-qubit coupling and the
resonator-mediated indirect coupling, and accordingly, the net coupling strength can be approximated by \cite{Yan2018}
\begin{equation}
\begin{aligned}\label{eq5}
&J_{xy}=J_{direct}+J_{indirect},
\\&J_{direct}=g_{12},
\\&J_{indirect}=\frac{g_{r1}g_{r2}}{2}(\frac{1}{\Delta_{1}}+\frac{1}{\Delta_{2}}),
\end{aligned}
\end{equation}
where $\Delta_{q}=\omega_{q}-\omega_{r}$ denotes the qubit-resonator detuning. The last term
is $ZZ$ coupling, which results mainly from the interactions among two-excitation subspace spanned
by $\{|101\rangle,|200\rangle,|020\rangle,|002\rangle\}$ (hereafter, the two-qubit system state
is denoted by the notation $|Q_{0},R,Q_{1}\rangle$) and is defined as \cite{DiCarlo2009,Galiautdinov2012}
\begin{equation}
\begin{aligned}\label{eq6}
\zeta_{zz}=(E_{101}-E_{100})-(E_{001}-E_{000}),
\end{aligned}
\end{equation}
where $E_{mnl}$ denotes the eigenenergy of the static Hamiltonian Eq.~(\ref{eq1})
with eigenstate $|\overline{mnl}\rangle$ and $m,n,l=\{0,1\}$. Note that qubit
operations, such as gate operations and measurements, are commonly performed in terms of the eigenbasis
of static (idle) Hamiltonians (see, e.g., Ref.~\cite{Galiautdinov2012}), hereinafter, our analysis
thus takes this into consideration implicitly.

As the direct qubit-qubit coupling in general is far smaller than the qubit-resonator coupling, the $ZZ$
coupling strength can be approximated as \cite{Zhao2021,Jin2021}
\begin{equation}
\begin{aligned}\label{eq7}
&\zeta_{zz}\simeq\zeta_{200}+\zeta_{002}+\zeta_{020},
\\&\zeta_{200}=\frac{-J_{002}^{2}}{\Delta+\alpha_{1}}\, {\rm with}\, J_{002}\simeq\sqrt{2}J_{xy},
\\&\zeta_{020}=\frac{J_{200}^{2}}{\Delta-\alpha_{2}}\, {\rm with}\, J_{200}\simeq\sqrt{2}J_{xy},
\\&\zeta_{002}=\frac{J_{020}^{2}}{\Delta_{1}+\Delta_{2}}\, {\rm with}\, J_{020}\simeq\sqrt{2}J_{indirect},
\end{aligned}
\end{equation}
where $\Delta=\omega_{1}-\omega_{2}$ is the qubit detuning and each term
comes from the interaction between the qubit state $|101\rangle$ and the high-energy
states $\{|200\rangle,|020\rangle,|002\rangle\}$ with the corresponding coupling
strengths $\{J_{200},J_{020},J_{002}\}$.

\subsubsection*{CR gate speed and ZZ crosstalk }\label{SecIIA2}

Here we now turn to briefly review the CR gate scheme for the coupled transmons discussed above (for more details, we
refer the reader to Refs.$\,$\cite{Malekakhlagh2020,Tripathi2019,Magesan2020}). The CR gate is implemented by applying a
microwave drive, i.e., CR drive, to the control qubit at the target qubit frequency. For the two-qubit system described by
the Hamiltonian Eq.~(\ref{eq1}), we consider that $Q_{1}$ is the control qubit while $Q_{2}$ is
the target, thus giving the following drive Hamiltonian
\begin{equation}
\begin{aligned}\label{eq8}
H_{d}=\Omega_{d}(t){\rm cos}(\omega_{d}t+\phi_{0})(a_{1}+a_{1}^{\dagger}),
\end{aligned}
\end{equation}
where $\omega_{d}$ is the drive frequency, and $\Omega_{d}(t)$ and $\phi_{0}$ are the drive amplitude
and initial phase, respectively. For simplicity, hereafter, we have $\phi_{0}=0$. Before going into details
of CR gates based on Eq.~(\ref{eq8}), we consider a more simple drive model based on
the effective two-qubit system described by Eq.~(\ref{eq4}), giving
\begin{equation}
\begin{aligned}\label{eq9}
\tilde{H}_{d}=\Omega_{d}(t){\rm cos}(\omega_{d}t)XI.
\end{aligned}
\end{equation}
As we will show, this model allows for a concise picture of the relation
between CR gate speeds and $ZZ$ crosstalk, and by combining with the following analysis
based on Eqs.~(\ref{eq1}) and~(\ref{eq8}), this model will provide a direct insight into the
effect of the qubit high-energy levels on CR gates.

By diagonalizing the effective Hamiltonian in Eq.~(\ref{eq4}) and then rewriting the drive
Hamiltonian Eq.~(\ref{eq9}) in the eigenstates, the full system Hamiltonian can be approximated by \cite{Chow2011}
\begin{equation}
\begin{aligned}\label{eq10}
&\bar{H}_{full}=\bar{\omega}_{1}\frac{ZI}{2}+\bar{\omega}_{2}\frac{IZ}{2}+\zeta_{zz}\frac{ZZ}{4}+\bar{H}_{d},
\\&\bar{H}_{d}=\Omega_{d}(t){\rm cos}(\omega_{d}t)[XI+\frac{J_{xy}}{\tilde{\Delta}}ZX],
\end{aligned}
\end{equation}
where $\tilde{\Delta}=\tilde{\omega}_{1}-\tilde{\omega}_{2}$ is the dressed detuning.
Here, the $XI$ term denotes the off-resonance drive on the control qubit, mainly
shifting the qubit frequency and thus finally contributing to a $ZI$ term, while
the $ZX$ term is the core of the gate scheme, enabling CR (ZX) gates with gate speeds
determined by the prefactor $\Omega_{d}J_{xy}/\tilde{\Delta}$. Note that to
mitigate the off-resonance error on the control qubit \cite{Malekakhlagh2022}, the drive amplitude is
generally smaller than the qubit detuning, thus the gate speed can be limited
by the interqubit coupling $J_{xy}$.

In Eq.~(\ref{eq10}), as the $ZZ$ term does not commute with the $ZX$ term, it can
degrade CR gates. To be more specific, for the typical CR architecture with capacitor
or resonator couplers (see Table~\ref{tab1:CR architectures}),
Eqs.~(\ref{eq5}) and~(\ref{eq7}) suggest that the $ZZ$ coupling strength is typically by
an order smaller than the $XY$ coupling strength. And to mitigate the error induced by
the $XI$ term in Eq.~(\ref{eq10}), the gate speed should be below $J_{xy}$. Considering these
findings, typically, an $XY$ coupling of $1.25\,\rm MHz$ can allow for a $200\,\rm-ns$ CX
gate and meanwhile leads to a residual $ZZ$ coupling of $\sim125\,\rm kHz$, which
contributes a gate error of $\sim 0.004$ (note that this is only a rough estimation
without considering the effect of high-energy levels on gate speeds) \cite{Kandala2021}. Moreover, the $ZZ$ term
can also cause idle errors and degrade individual qubit addressability \cite{Gambetta2012a,McKay2019}. In
this sense, the ZZ coupling acts as quantum crosstalk degrading the operational fidelity in CR architectures.
Indeed, according to Eq.~(\ref{eq7}), while the crosstalk can be suppressed by decreasing the
interqubit coupling, this, in turn, slows down the CR gate speed. In essence, this
means that there exists a speed-fidelity trade-off imposed by the $ZZ$ crosstalk.

In the above discussion, while the effect of the high-energy levels has been taken into account
in analyzing $ZZ$ crosstalk, their effect on the CR gate speed has not been explored. Here, based
on the drive Hamiltonian Eq.~(\ref{eq8}), we summarize the main results on this
subject. To simplify the analysis, we consider that two transmons are coupled directly via a capacitor
coupler, as indicated in Table~\ref{tab1:CR architectures}, and the static system Hamiltonian
is given by Eq.~(\ref{eq1}) with $g_{rq}=0$. Again, by rewriting the drive Hamiltonian
in the eigenstates of the static Hamiltonian and then truncating to the qubit subspace, the drive
Hamiltonian has the following approximate form
\begin{equation}
\begin{aligned}\label{eq11}
\bar{H}_{d}=\Omega_{d}(t){\rm cos}(\omega_{d}t)[XI+\mu_{ix}IX+\mu_{zx}ZX],
\end{aligned}
\end{equation}
where the prefactors of the $IX$ and $ZX$ terms are
\begin{equation}
\begin{aligned}\label{eq12}
&\mu_{ix}=\langle \overline{101}|(a_{1}+a_{1}^{\dagger})|\overline{100}\rangle+\langle \overline{001}|(a_{1}+a_{1}^{\dagger})|\overline{000}\rangle,
\\&\mu_{zx}=\langle \overline{101}|(a_{1}+a_{1}^{\dagger})|\overline{100}\rangle-\langle \overline{001}|(a_{1}+a_{1}^{\dagger})|\overline{000}\rangle.
\end{aligned}
\end{equation}
By expanding to the first order in $J_{xy}/\Delta$, the above expressions can be approximated by \cite{Malekakhlagh2020,Tripathi2019,Magesan2020}
\begin{equation}
\begin{aligned}\label{eq13}
&\mu_{ix}=\frac{-J_{xy}\Delta}{\Delta(\alpha_{1}+\Delta)},\,\mu_{zx}=\frac{J_{xy}\alpha_{1}}{\Delta(\alpha_{1}+\Delta)}.
\end{aligned}
\end{equation}

From Eqs.~(\ref{eq10}-\ref{eq13}), different from the result based on the two-level qubit model, there exists an
additional term $IX$, which commutes with the $ZX$ term and thus does not degrade the CR gate and its speed.
Importantly, assuming fixed $J_{xy}$, for positive control-target detunings, $\mu_{zx}$ can have
a larger magnitude than that with negative detunings, due to the negative transmon anharmonicity. As a result, to
ensure fast gates, the positive detuning is favored while the negative detuning is identified as the slow
gate condition \cite{Brink2018,Hertzberg2021}. Besides, from Eq.~(\ref{eq13}), $\mu_{zx}$ can be suppressed
with large detunings. Thus, the straddling regime, i.e., $0<\Delta<-\alpha_{1}$ \cite{Koch2007}, is
often preferred for fast-speed CR gates. Note that assuming an infinite anharmonicity $\alpha_{1}\rightarrow\infty$, the above
expression reduces to Eq.~(\ref{eq10}), confirming that the main difference between
the result based on the two-level qubit model in Eq.~(\ref{eq4}) and the result shown here
comes from the qubit high-energy levels.

\subsubsection*{Frequency collisions and Frequency allocation}\label{SecIIA3}

The above discussions mainly focus on the most basic unit, which comprises two fixed-frequency transmons with
fixed coupling, of CR-based transmon architectures. Here, we briefly discuss one of the most fundamental
challenges in scaling up this architecture, i.e., frequency collisions. To start, we consider a
fixed-coupled system of three transmons, labeled by $Q_{i}$, $Q_{j}$, and $Q_{k}$, as shown in
Fig.~\ref{fig1}(a), where $Q_{j}$ is the control and $Q_{i}$ and $Q_{k}$ are targets. We further
assume that the typical interqubit coupling and the detuning for control-target qubit pairs
are $J$ and $\Delta$, respectively, and the typical CR drive magnitude is $\Omega$. As mentioned
above, to ensure fast gates, the qubit system is operated with positive control-target
detunings ($\omega_{j}>\omega_{i(k)}$) in the straddling regime.

Generally, frequency collisions can be classified into two major types: \emph{the static frequency
collision} due to always-on couplings and \emph{the dynamic frequency collision} related to CR
drive-induced transitions. As in existing literature \cite{Brink2018,Hertzberg2021}, here we list six main
types of frequency collisions, which are caused by qubit-qubit couplings
with strengths of $\sim J$ and by transitions with rates of $\sim J\Omega/\Delta$
or $\sim \Omega^{2}/\Delta$ (to achieve high-fidelity gates, these couplings and
transitions give larger collision bounds than that arising from even higher-order
couplings and transitions with strengths of, e.g., $J^{2}/\Delta$ and $J^{2}\Omega^{2}/\Delta^{3}$,
which will be discussed in Sec.~\ref{SecIII}):

(1) \emph{The static frequency collision}. Because of the always-on couplings: (i) for the
static systems, any degeneracy in qubit state transitions, such
as $|0\rangle\leftrightarrow|1\rangle$, can lead to a strong state hybridization
between qubits, degrading individual qubit addressability; (ii) for systems
under CR drives, off-resonance drive can shift qubit frequencies and can cause qubits
on-resonance with others, resulting in population swaps between qubits. For neighboring
qubits with coupling strengths of $\sim J$, there exist two most likely
cases, (S1) $\omega_{j}=\omega_{i(k)}$ and (S2) $\omega_{j}=\omega_{i(k)}+\alpha_{i(k)}$.

(2) \emph{The dynamic frequency collision}. This arises from unwanted transitions
activated by CR drives and thus both the transition rates and the collision
conditions (i.e., considering that the drive itself can stark-shift system energy levels) depend
on the drive. For a CR gate applied to $Q_{j}$ and $Q_{i}$, a drive
at $Q_{i}$'s frequency should be applied to $Q_{j}$. For $Q_{j}$ itself, when the
drive is on-resonance with its transitions, this will lead to three collisions and
two of the three also cause the static collisions, i.e., (S1) and (S2) (note that
depends on the specific (static or dynamic) situation, ensuing high-fidelity gates
will give different bounds on $\Delta$). The
newly added one is (D1) the two-photon transition $|0\rangle\leftrightarrow|2\rangle$
with the condition of $\omega_{i(k)}=\omega_{j}+\alpha_{j}/2$. Besides, similar to $Q_{i}$, the drive can
also be felt by the neighboring spectator $Q_{k}$, due
to the always-on coupling between $Q_{j}$ and $Q_{k}$. Thus, when $Q_{i}$'s frequency, i.e., the
CR drive frequency, is on-resonance with any transitions
of $Q_{k}$, this can cause unwanted transitions of $Q_{k}$. This leads to two
collision cases, i.e., (D2) the transition $|0\rangle\leftrightarrow|1\rangle$
with $\omega_{i}=\omega_{k}$, (D3) the transition $|1\rangle\leftrightarrow|2\rangle$
with $\omega_{i}=\omega_{k}+\alpha_{k}\,(\omega_{k}=\omega_{i}+\alpha_{i})$. Finally,
there exists another condition (D4) $2\omega_{j}+\alpha_{j}=\omega_{i}+\omega_{k}$, which
corresponds to the drive-assisted transition $|100\rangle(|001\rangle)\leftrightarrow|020\rangle$.

From the above discussion, one can find that similar to the $ZZ$ crosstalk, these
frequency-collision issues mainly arise from the always-on interqubit couplings and the
weak anharmonicity of transmons. More importantly, most of them can act as quantum crosstalk, limiting gate performance
and imposing speed-fidelity tradeoffs. Thus, to ensure fast, high-fidelity operations
for multiqubit lattices, these frequency-collision conditions should be avoided, further reducing
the usable frequency range determined by the weak transmon anharmonicity. For example, assuming a
typical $250\rm ns$ CR gate, the qubit-qubit couplings and the rates of the drive-induced transitions
are generally of the order of $1\,\rm MHz$, thus to ensure gate error of $\sim 0.01$ ($0.001$), the
qubit system should be operated outside of the bound of about $\pm 10\,\rm MHz$ ($\pm 30\,\rm MHz$) around
each frequency collision. 

To this end, scaling up the CR-based transmon architectures requires a
minimum of five distinct frequencies, i.e., $\omega_{5}>\omega_{4}>\omega_{3}>\omega_{2}>\omega_{1}$, for square qubit
lattices \cite{Brink2018,Hertzberg2021,Chamberland2020}, while for heavy-hexagonal qubit lattices, as shown
in Fig.~\ref{fig1}(a), a minimum number of three, i.e., $\omega_{3}>\omega_{2}>\omega_{1}$, is
required \cite{Hertzberg2021,Chamberland2020}. However, considering fabrication uncertainty, whether such
frequency allocation can successfully mitigate frequency collisions depends on how precise in setting
qubit frequencies is, as shown, for example, in Fig.~\ref{fig1}(d). Recent studies show
that, for a heavy-hexagonal qubit lattice, the state-of-the-art precision in setting frequencies allows
the CR architecture to scaling up to the 100-qubit size while roughly a factor of two further
improvement is needed for 1000-qubit size \cite{Hertzberg2021}. Overall, these findings suggest that
frequency collision is still the major challenge for scaling up the
CR-based transmon architecture.

\begin{figure*}[tbp]
\begin{center}
\includegraphics[width=18cm,height=6cm]{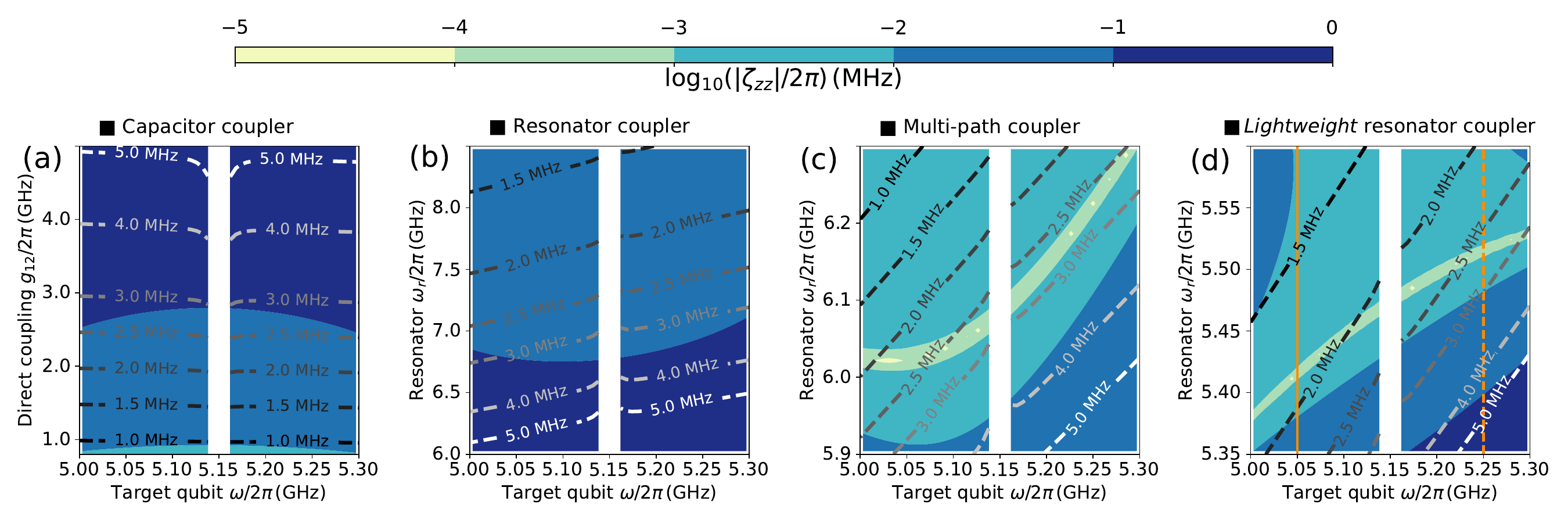}
\end{center}
\caption{Landscapes of ZZ coupling $\zeta_{zz}$ (the filled contours) and XY coupling $J_{xy}$ (the dashed contours).
Here, the control qubit frequency is $\omega_{1}/2\pi=5.15\,\rm GHz$, and the coupling parameters and the qubit anharmonicity
are listed in Table~\ref{tab1:CR architectures}. (a) For qubits coupled via capacitor couplers, the contours show
the $ZZ$ and $XY$ coupling strengths as functions of the target qubit frequency and the direct coupling strength.
(b), (c), and (d) show the $ZZ$ and $XY$ coupling strengths as functions of the target qubit frequency and the
resonator frequency, for qubits coupled via resonator couplers, multipath couplers, and the introduced lightweight
resonator couplers, respectively. Vertical cuts, indicated by the solid and dashed orange lines, denote the result
plotted in Figs.~\ref{fig3}(a) and~\ref{fig3}(b), respectively.}
\label{fig2}
\end{figure*}

\subsection{CR architecture with passive mitigation of ZZ crosstalk and frequency collisions}\label{SecIIB}

As discussed above, in the context of CR-based transmon architecture, quantum crosstalk
and frequency collisions are two major challenges to be faced when scaling up this
architecture. In the following discussion, we will show how to passively mitigate both
of two issues by introducing a promising CR architecture based on lightweight
resonator couplers. Additionally, note that as mentioned before, the straddling
regime is preferred to ensure fast-speed CR gates. Thus, hereafter, unless otherwise
specified, we always consider this implicitly.

\subsubsection*{System Hamiltonian and ZZ suppression}\label{SecIIB1}

As mentioned in Sec.~\ref{SecI}, in our proposed CR architecture, the fixed-frequency
transmon qubits are coupled via a lightweight resonator, and the typical coupling parameters
are tabulated in the last column of Table~\ref{tab1:CR architectures}. For two transmons coupled via
such a coupler, the static system Hamiltonian can be described by Eq.~(\ref{eq1}) with $g_{12}=0$. After
applying the rotating-wave approximation (RWA), the Hamiltonian is given by
\begin{equation}
\begin{aligned}\label{eq14}
H_{static}=&\sum_{q=1,2}\big(\omega_{q}a_{q}^{\dagger}a_{q}+\frac{\alpha_{q}}{2}a_{q}^{\dagger}a_{q}^{\dagger}a_{q}a_{q}\big)+\omega_{r}a_{r}^{\dagger}a_{r}
\\&+\sum_{q}g_{rq}(a_{r}a_{q}^{\dagger}+a_{r}^{\dagger}a_{q}).
\end{aligned}
\end{equation}
As in Sec.~\ref{SecIIA}, truncated to the qubit subspace, the above Hamiltonian can be
approximated by Eq.~(\ref{eq4}) with the $XY$ and the $ZZ$ coupling strengths given as \cite{DiCarlo2009}
\begin{equation}
\begin{aligned}\label{eq15}
&J_{xy}=J_{indirect}=\frac{g_{r1}g_{r2}}{2}(\frac{1}{\Delta_{1}}+\frac{1}{\Delta_{2}}),
\\&\zeta_{zz}=2J_{xy}^{2}\bigg[\frac{1}{\Delta-\alpha_{2}}
-\frac{1}{\Delta+\alpha_{1}}+\frac{1}{\Delta_{1}+\Delta_{2}}\bigg].
\end{aligned}
\end{equation}
Note that as already mentioned in Sec.~\ref{SecIIA}, stray couplings between
qubits are ubiquitous in real devices. In Appendix~\ref{A}, we provide specific
cases for illustrating their effects on the current studied architecture and show that
their presence does not change the main results based
on Eq.~(\ref{eq14}). Accordingly, in the following discussion, to avoid
repetition, we will focus only on the qubit architecture described
by Eq.~(\ref{eq14}), in which the direct qubit-qubit coupling is not included.

From Eq.~(\ref{eq15}), there are two essential features in the current architecture: (1) a $ZZ$-free point can exist
due to the destructive interference of ZZ coupling contributing
from interactions between computational state $|101\rangle$ and high-energy
states $\{|200\rangle,|002\rangle,|020\rangle\}$ \cite{Jin2021,Li2022,Zhao2022a} and (2) its existence
does not depend on the$XY$ coupling strength $J_{xy}$. The two features allow us to suppress $ZZ$ coupling
while retaining $XY$ coupling. For instance, considering that when the qubit
detuning is far smaller than the qubit anharmonicity (i.e., $|\Delta|\ll |\alpha_{q}|$), according
to Eq.~(\ref{eq15}), the ZZ-free point exists when the qubit-resonator detuning is comparable to
the qubit anharmonicity, i.e., $\Delta_{q}\simeq \alpha_{q}$. Accordingly, the strength of the
maintained $XY$ coupling is about $g_{r1}g_{r2}/\alpha_{q}$. For the typical transmon
anharmonicity of $-330\,\rm MHz$ and within the dispersive regime, i.e, $|\Delta_{q}/g_{rq}|\gtrsim 10$, $XY$
couplings with the strengths of $1\mbox{-}3\,\rm MHz$ can be obtained, potentially allowing
fast-speed CR gates. By contrast, according to Eqs.~(\ref{eq7}) and~(\ref{eq15}), the
$ZZ$ expression for direct-coupled qubits is then expressed as \cite{Barends2014}
\begin{equation}
\begin{aligned}\label{eq16}
\zeta_{zz}=2J_{xy}^{2}\bigg[\frac{1}{\Delta-\alpha_{2}}-\frac{1}{\Delta+\alpha_{1}}\bigg]\,{\rm with}\, J_{xy}=g_{01},
\end{aligned}
\end{equation}
which illustrates that no $ZZ$-free point exists in the direct-coupled architectures.
Meanwhile, for the architecture with resonator couplers shown in Table~\ref{tab1:CR architectures},
when one allows this architecture to be operated outside of the dispersive
regime, i.e., quasi-dispersive regime, $|\Delta_{q}/g_{rq}|< 10$, the $ZZ$ coupling can also be suppressed
while maintaining even larger $XY$ couplings \cite{Goerz2017,Li2022}. However, in this situation, besides
that $ZZ$ suppression depends strongly on system parameters (e.g., is very sensitive to small drifts in qubit and
resonator frequencies), the strong state hybridization between qubits and the resonator will degrade individual qubit
addressability and aggravate the frequency-collision issues.

We now turn to give a more quantitative analysis of the introduced architecture. Moreover, for ease of
comparison, we also show the results for the existing architectures tabulated in
Table~\ref{tab1:CR architectures}. Within the CR architecture,
Figure~\ref{fig2} shows contour plots of the $ZZ$ (the filled contours) and $XY$ (the dashed contours) coupling
strengths as functions of system parameters, with the control qubit frequency of $5.15\,\rm GHz$
and the other qubit parameters listed in Table~\ref{tab1:CR architectures}. Here, according
to Eq.~(\ref{eq6}), $ZZ$ coupling strengths are extracted by numerical diagonalization of the
static system Hamiltonian, while similar to Eqs.~(\ref{eq12}) and~(\ref{eq13}), $XY$ coupling strengths are
inferred from the following approximation \cite{Tripathi2019}
\begin{equation}
\begin{aligned}\label{eq17}
-\frac{J_{xy}}{\Delta}\simeq\langle \overline{001}|(a_{1}+a_{1}^{\dagger})|\overline{000}\rangle .
\end{aligned}
\end{equation}
To verify this, Figure~\ref{fig2}(a) shows both the inferred $J_{xy}$ values and the setting
values, $J_{xy}=g_{12}$. One can find that the inferred values mostly agree well with the
setting values. This justifies the estimation of $J_{xy}$ based on Eq.~(\ref{eq17}).

\begin{figure}[tbp]
\begin{center}
\includegraphics[keepaspectratio=true,width=\columnwidth]{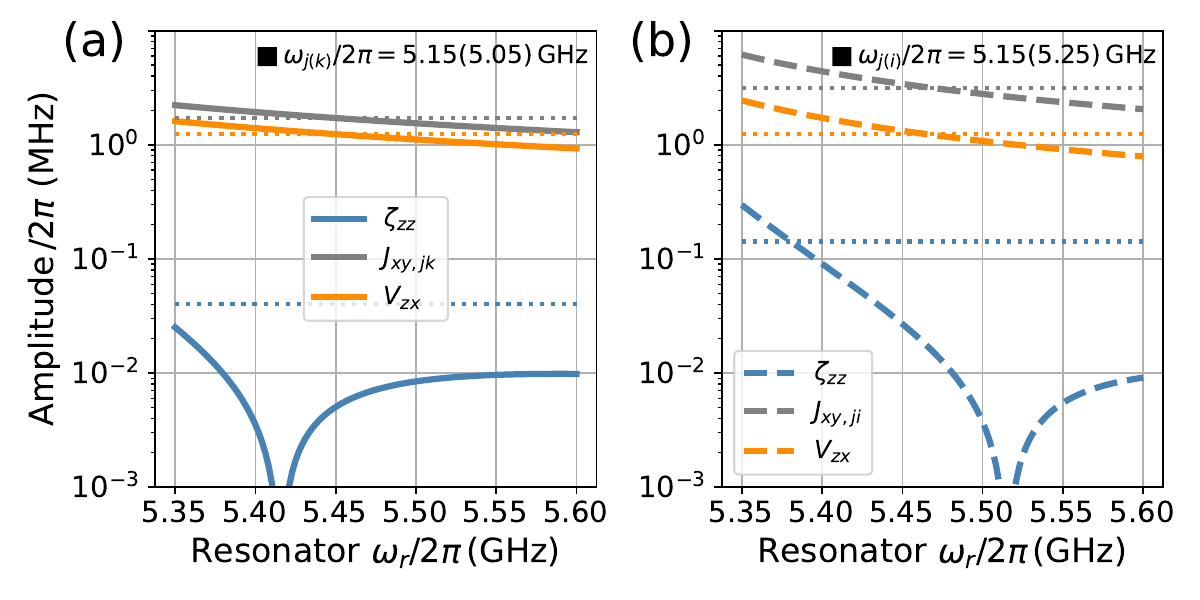}
\end{center}
\caption{The magnitudes of the $ZZ$ coupling strength $\zeta_{zz}$, the $XY$ coupling strength $J_{xy}$, and
the $ZX$ rate $V_{zx}$ (inferred by assuming a CR drive magnitude of $50\,\rm MHz$) as functions of
the resonator frequency in the proposed architecture. Here, plots of $\zeta_{zz}$ and $J_{xy}$ denote
two one-dimensional vertical cuts through Fig.~\ref{fig2}(d) (indicated by the solid and dashed orange lines) at
the target qubit frequencies of (a) $\omega_{k}/2\pi=5.05\,\rm GHz$
and (b) $\omega_{i}/2\pi=5.25\,\rm GHz$, respectively, and the control qubit frequency is $\omega_{j}/2\pi=5.15\,\rm GHz$.
The dotted lines represent the results for direct-coupled qubits with the target $ZX$ rate
of $1.25\,\rm MHz$, allowing $200\,\rm-ns$ CX gates without rise-fall times.}
\label{fig3}
\end{figure}

From the results shown in Figs.~\ref{fig2}(a) and~\ref{fig2}(b), as expected, in
architectures with capacitor or resonator couplers, no ZZ-free points exist and the $ZZ$ coupling strength
is typically ten-fold smaller than that of the $XY$ coupling, giving rise to $|J_{xy}/\zeta_{zz}|\sim 10$.
However, as shown in Fig.~\ref{fig2}(d), for the lightweight resonator coupler, ZZ-free points exist
and the $ZZ$ coupling can be heavily suppressed in a wide frequency range while the maintained $XY$ coupling
strength typically ranges from $1\,\rm MHz$ to $3\,\rm MHz$. Remarkably, compared to the capacitor and resonator couplers, this
give rise to $|J_{xy}/\zeta_{zz}|> 10^{2}$, providing at least one order-of-magnitude improvement in $ZZ$ suppression.
More specifically, vertical cuts through Fig.~\ref{fig2}(d) at
the target qubit frequencies of $5.05\,\rm GHz$ and $5.25\,\rm GHz$ are shown in Fig.~\ref{fig3}, where dotted
lines denote the results for direct-coupled qubits with the target of ensuring $200\,\rm-ns$ CX gates (assuming the CR drive
amplitude of $50\,\rm MHz$ and ignoring rise-fall times). In contrast to the direct-coupled case, where the
residual $ZZ$ coupling strength is typically in the range of $50\mbox{-}100\,\rm kHz$, the lightweight resonator coupler
can allow the $ZZ$ suppression to below $10\,\rm kHz$, in both the positive [see Fig.~\ref{fig3}(a)] and
negative [see Fig.~\ref{fig3}(b)] control-target detuning regions.

As shown in Fig.~\ref{fig2}(c) and demonstrated in the previous works \cite{Kandala2021,Zhao2021}, similar
improvements can also be available for architectures with multipath couplers. In particular, as the design
of multipath couplers generally increases the circuit complexity, so their success on $ZZ$ suppression is potentially
more prone to imperfections attributable to the design itself and fabrications. Meanwhile, the result shown in
Fig.~\ref{fig2}(d) demonstrates that even with a simpler coupling circuit, $ZZ$ suppression could still be
achieved while maintaining large $XY$ couplings. However, note that the lightweight resonator coupler
also has its own drawbacks as well, which will be discussed in Sec.~\ref{SecIV}.

\subsubsection*{Revisiting the slow gate region}\label{SecIIB2}

The above discussions have illustrated that with lightweight resonator couplers, $ZZ$ crosstalk can be
mitigated while maintaining large $XY$ couplings. More importantly, in contrast to the cases with
capacitor or resonator couplers [see Figs.~\ref{fig2}(a) and~\ref{fig2}(b)], here one can find a remarkably
strong dependence of the $XY$ coupling strength $J_{xy}$ on the control-target detuning, as shown
in Fig.~\ref{fig2}(d). This can be more specific, in that
as illustrated in Fig.~\ref{fig3}, compared to that in positive detuning regions, a typical
two-fold increase in $J_{xy}$ can be obtained without affecting the $ZZ$ suppression within
the negative detuning regions. As expected from Eq.~(\ref{eq13}), the increased $XY$ coupling
can compensate for the loss in gate speed due to the negative transmon anharmonicity. Therefore, fast-speed
CR gates could also be available within the negative detuning regions, which have previously been 
identified as forbidden zones with slower gate speeds \cite{Brink2018,Hertzberg2021}.

The origin of the above appealing features can be rationalized by noting that
from Eq.~(\ref{eq15}), when the qubit-resonator detuning is comparable to the qubit-qubit detuning (as in
the currently studied architecture), i.e., $|\Delta_{rq}|\sim |\Delta|$, $J_{xy}$ will depend
strongly on the detuning sign. By assuming a fixed control qubit frequency, having negative
control-target detunings means that the detuning between the resonator and the target qubit is smaller
than that with positive detunings, enabling larger resonator-mediated $XY$ couplings. By
contrast, if $|\Delta_{rq}|\gg|\Delta|$ (as in the architecture with resonator
couplers), whether the detuning changes its sign, there is little effect on $J_{xy}$, as shown in
Fig.~\ref{fig2}(b). Therefore, in architectures with capacitor and resonator couplers, to increase $XY$
coupling, one has to increase the direct coupling or the qubit-resonator coupling and this will inevitably
increase $ZZ$ couplings. For example, under the conditions with negative detunings, to ensure gate
speed comparable to that in the positive detuning regions, the direct coupling strength should be increased
from $1.7\,\rm MHz$ to $3.1\,\rm MHz$ and accordingly, the $ZZ$ coupling will increase
from $41\,\rm kHz$ to $142\,\rm kHz$, as shown by the dotted lines in Fig.~\ref{fig3}.
This further explicitly explains why the negative detuning is previously identified as the slow
gate condition \cite{Brink2018,Hertzberg2021}. On the contrary, here all the benefits
from the increased $XY$ coupling can be achieved without
sacrificing $ZZ$ suppression. Additionally, Figure~\ref{fig2}(c) also shows similar appealing
behavior, suggesting that the multipath coupler can be an alternative for achieving fast gate speed
in negative detuning regions.

\begin{figure}[tbp]
\begin{center}
\includegraphics[keepaspectratio=true,width=\columnwidth]{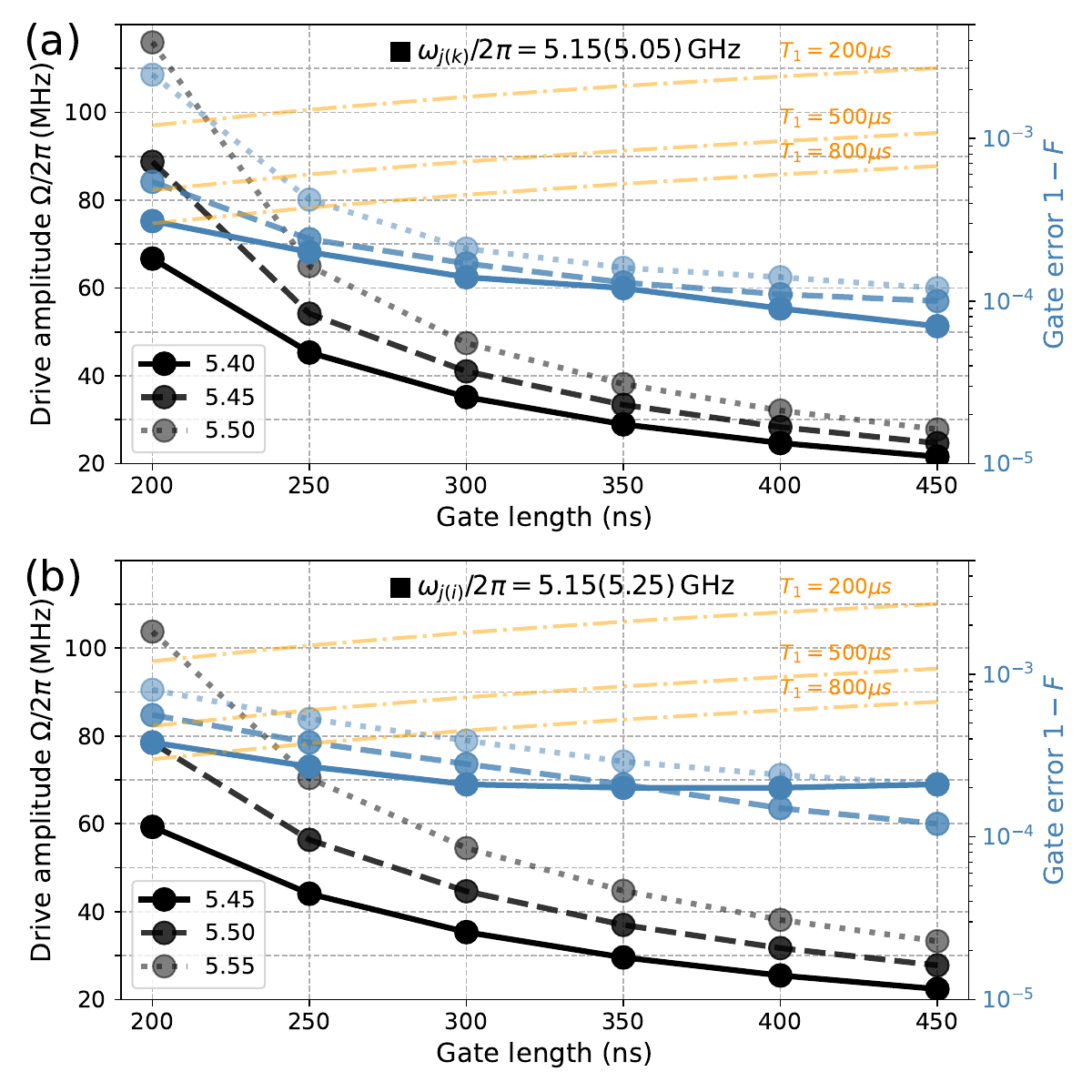}
\end{center}
\caption{CX gate errors (in absence of decoherence) and peak drive amplitudes as functions of the gate length with varying
resonator frequencies. Here, the control qubit frequency is $\omega_{j}/2\pi=5.15\,\rm GHz$, and the coupling
parameters and the qubit anharmonicity are listed in Table~\ref{tab1:CR architectures}. Orange
dash-dotted lines denote the incoherence error $\epsilon_{incoh}$ versus the gate length (the coherence times are assumed to be the same for both the control
and target qubits and $T_{2}=T_{1}$, with different $T_{1}$).(a) The target qubit frequency
is $\omega_{k}/2\pi=5.05\,\rm GHz$, giving rise to the condition of positive control-target
detunings. (b) The target qubit frequency is $\omega_{k}/2\pi=5.25\,\rm GHz$, giving rise to the conditions of negative control-target
detunings.}
\label{fig4}
\end{figure}

In particular, Figure~\ref{fig3} also shows the $ZX$ rate $V_{zx}$ as
a function of the resonator frequency. Here, according to Eq.~(\ref{eq12}), $V_{zx}$ is determined by
\begin{equation}
\begin{aligned}\label{eq18}
V_{zx}=\frac{\Omega}{2}\mu_{zx}.
\end{aligned}
\end{equation}
with the assumption of the CR drive magnitude $\Omega/2\pi=50\,\rm MHz$. As expected, even under
the condition of negative detunings, the resulting $ZX$ rate is comparable to that within the positive
detuning regions. While here the typical $ZX$ rate is about $1\,\rm MHz$, allowing $250\,\rm-ns$ CX
gates without rise-fall times, even higher $ZX$ rates may be possible by increasing the
qubit-resonator coupling $g_{rq}$. However, increasing $g_{rq}$ generally will increase the
residual $ZZ$ couplings. For example, when increasing the qubit-resonator coupling
from $25\,\rm MHz$ to $30\,\rm MHz$, the typical $ZX$ rate becomes $1.5\,\rm MHz$ while the
typical $ZZ$ coupling has increased to $20\,\rm kHz$ (see Appendix~\ref{B} for details).
Given the state-of-the-art gate performance is still limited by qubit decoherence, presently unitizing
larger qubit-resonator couplings could help increase the gate speed and reduce
the incoherent error. This highlights the importance of balancing incoherent errors and coherence
errors arising, for example, from $ZZ$ crosstalk, to achieve high-fidelity gates.

Below, to provide a more direct verification, we turn to study the implementation of $CX$ gates within
the introduced architecture. In particular, we will focus on both the conditions of positive and negative
control-target detunings. Here, based on the CR gate scheme, we consider the realization of direct CX
gates \cite{Jurcevic2021,Malekakhlagh2022}. Besides the CR drive on the control qubit, here, an additional cancelation drive
is also applied to the target qubit at its frequency. This additional drive is introduced to ensure no operation
on the target qubit when the control qubit is in state $|0\rangle$ \cite{Malekakhlagh2022}. Accordingly, within the RWA, the
drive Hamiltonian can be expressed as
\begin{equation}
\begin{aligned}\label{eq19}
H_{drive}=\sum_{q=1,2}\frac{\Omega_{q}(t)}{2}(a_{1}^{\dagger}e^{-i\omega_{d}t}+a_{1}e^{i\omega_{d}t}),
\end{aligned}
\end{equation}
where $\Omega_{q}(t)$ denotes the amplitude of the drive on $Q_{q}$. Hereafter, we consider using
the cosine-decorated square pulse with fixed ramp times of $20\,\rm ns$, which is defined as
\begin{align}
\Omega_{d}(t)\equiv
\begin{cases}
\Omega\frac{1-\cos{(\pi \frac{t}{t_r}})}{2}  \;, &0<t<t_r\\
\Omega\;,  &t_r<t<t_g-t_r\\
\Omega\frac{1-\cos{(\pi \frac{t_g-t}{t_r}})}{2} \;, &t_g-t_r<t<t_g
\end{cases}
\label{eq20}
\end{align}
where $\Omega$ denotes the peak drive amplitude, $t_r=20\,\rm ns$ is the ramp time, and $t_g$ represents
the gate length. A detailed procedure on the gate tune-up and the characterization for the
CX gate can be found in Appendix~\ref{C}.

Figure~\ref{fig4} shows both the CX gate error (intrinsic error, i.e., excluding the gate error from qubit incoherence) \cite{Pedersen2007} and
the CR drive amplitude as functions of the gate length with varying resonator frequencies. As expected
from the above analysis, under the conditions of negative detunings, high-fidelity CX gates can still
be achieved with a gate speed comparable to that within the positive detuning regions. Moreover, the
results shown in Fig.~\ref{fig4} also illustrate that within the typical range of $20\mbox{-}70\,\rm MHz$
drive amplitude, $CX$ gates with gate errors approaching $10^{-4}$ could be obtained with a gate
length of $250\mbox{-}450\,\rm ns$. In addition, as studied in previous
works \cite{Motzoi2009,Gambetta2011,Li2023,Malekakhlagh2022}, by optimizing
drive parameters, the pulse ramp time can be shorter and the off-resonance error due to the CR drive can be
further suppressed. Moreover, as mentioned
before, even higher $ZX$ rates (e.g., $1.5\,\rm MHz$, allowing for a $200\,\rm ns$ CX gate with a ramp
time of $20\,\rm ns$) can be obtained by increasing the qubit-resonator coupling. Meanwhile, the
resulting $ZZ$ coupling can be suppressed below $20\,\rm kHz$, contributing to the gate error at the
level of $10^{-4}$ \cite{Kandala2021}. Thus, CX gates with even faster gate speed and higher gate fidelity should be possible
within the currently studied CR architecture.

Furthermore, in Fig.~\ref{fig4}, we also show the
incoherence error versus the gate length for the CX gate. Here, the coherence times are assumed to be the same
for both the control and target qubits with $T_{2}=T_{1}$ \cite{Malekakhlagh2022,Kim2023}, and the incoherence error can
be estimated by $\epsilon_{\rm incoh}\approx\sum_{q=1,2}[(1/5)(t_{g}/T_{1}^{(q)})+(2/5)(t_{g}/T_{2}^{(q)})]$
\cite{Malekakhlagh2022}. One can clearly see from Fig.~\ref{fig4} that there exists
a trade-off between the intrinsic (control) gate error (mainly dominated by the control error from the off-resonance
CR drive) and the incoherence error. Given the state-of-the-art coherent times of fixed-frequency
qubits \cite{Place2021,Wang2022,Gordon2022}, the gate error should be limited
by the qubit incoherence in both the negative and positive detuning regimes.

As in various qubit architectures with superconducting qubits, gate speeds and operational performance
generally hinge on the ability in setting qubit parameters. Specifically, in the currently studied CR
architecture and for fixed resonator-qubit couplings (see Figs.~\ref{fig3} and~\ref{fig4}), the
gate speed is mainly determined by the qubit-resonator detuning. Accordingly, assuming the drive
amplitude of $50\,\rm MHz$, to ensure $300\,\rm-ns$ CX gates (with a ramp time of $20\,\rm ns$), the
usable range of the qubit-resonator detuning is about $200\,\rm MHz$ for the positive detuning case while
it reduces to about $100\,\rm MHz$ for the negative detuning case, as shown in
Figs.~\ref{fig3} and~\ref{fig4}. In contrast to the architecture with resonator
couplers, where the $XY$ coupling (thus also the CR gate speed) shows less sensitivity to the variation
in qubit-resonator detuning, see Figs.~\ref{fig2}(b), this generally introduces an additional constraint
for achieving fast-speed CR gates. However, as illustrated in Appendix~\ref{B}, increasing
resonator-qubit couplings could help relieve this constraint. Moreover, since currently the
reproducibility of the resonator frequency is much better than that of qubits (see, e.g., Ref.~\cite{Norris2023}), this
constraint could be largely addressed by improving the accuracy in setting qubit frequencies (as for
addressing the frequency-collision issue, but with less stringent requirements on accuracy).

Given the above illustration of fast-speed CX gates in the negative detuning region, we now go back
to the frequency-collision issue. As discussed in Sec.~\ref{SecIIA}, the qubit frequency allocation
should be optimized to balance gate errors due to frequency collisions and incoherence errors
resulting from slow gate speeds. Since currently, gate performance is most likely to be
limited by qubit decoherence, the positive detuning region is preferred for ensuring fast-speed
gates. However, this in turn decreases the usable range of qubit detunings, making the frequency-collision
issue more prominent. Thus, mitigating frequency collisions and improving the collision-free
yield of large-size systems put stringent requirements on the accuracy in
setting qubit frequencies \cite{Brink2018,Morvan2022,Hertzberg2021}. Even with sparse connectivity, e.g., the heavy-hexagonal lattice
topology shown in Fig.~\ref{fig1}(a), the state-of-the-art accuracy is still not enough to
support $1000$-qubit systems \cite{Hertzberg2021}. Within the current studied CR architecture, we show that fast-speed gates are also
available in the negative detuning regions, extending the usable detuning range. As a
consequence, we expect that this could largely mitigate the frequency-collision issue and relax the stringent
requirement on the accuracy for scaling up to large system sizes.

\section{Crosstalk mitigation in multiqubit systems}\label{SecIII}

\begin{figure}[tbp]
\begin{center}
\includegraphics[keepaspectratio=true,width=\columnwidth]{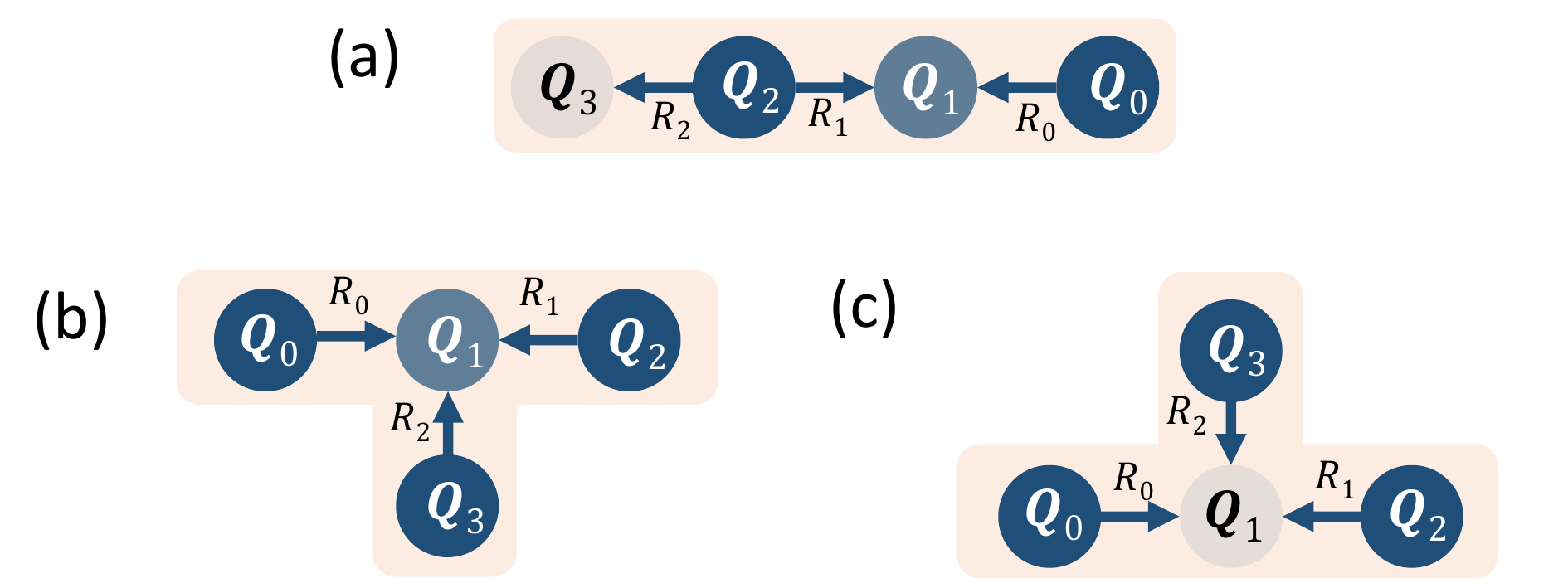}
\end{center}
\caption{Typical four-qubit units in the heavy-hexagonal qubit lattice
with the three-frequency pattern shown in Fig.~\ref{fig1}(a). The three
four-qubit systems,i.e., (a) '$-$'-shape, (b) '$\top$'-shape, and (c) '$\bot$'-shape, are
also shown in the yellow shadow of Fig.~\ref{fig1}(a) and here the
arrows indicate the direction of the CR gates, i.e., from control
to target.}
\label{fig5}
\end{figure}

In this section, to examine the practicality of the architecture, we intend to
investigate the CX gate performance and the impact of quantum crosstalk on it in the
typical four-qubit units of the heavy-hexagonal qubit lattice, as shown in Fig.~\ref{fig5}.
Then, we turn to give schemes mitigating the crosstalk effect on gates. As shown in
Fig.~\ref{fig1}(e), unlike in other existing literature [see Fig.~\ref{fig1}(d)], here the
three-frequency pattern is given as $\omega_{i}>\omega_{j}>\omega_{k}$. Following
the above discussion, here we consider that the desired control frequency
is $\omega_{j}/2\pi=5.15\,\rm GHz$ while the target qubit frequency
is $\omega_{i(k)}/2\pi=5.25(5.05)\,\rm GHz$. This frequency allocation scheme
is expected to mitigate the frequency-collision issue within the current
architecture, as illustrated in Figs.~\ref{fig1}(d) and~\ref{fig1}(e).

In the following discussion, we first study the gate performance of isolated CX gates
in the four-qubit systems and then turn to study the simultaneous gates. Moreover, we also
provide a preliminary study to determine the effect of frequency variations
on $ZZ$ suppression and gate performance.

\begin{table}[tbp!]
\caption{Typical qubit and resonator parameters used in studying the four-qubit units shown
in Fig.~\ref{fig5}. The other system parameters are listed in Table~\ref{tab1:CR architectures}
and the resulting $ZZ$ couplings range from $1\,\rm kHz$ and $13\,\rm kHz$ for near-neighbor
qubits. Here, the qubit and resonator frequencies are given in units of $\rm GHz$.\label{tab2:System parameters}}
\begin{ruledtabular}
\begin{tabular}{cccccccc}
   Layout & $Q_{0}$ & $R_{0}$ & $Q_{1}$ &$R_{1}$ & $Q_{2}$ &$R_{2}$ & $Q_{3}$ \\
   \hline
  '$-$'-shape & 5.15 & 5.40 & 5.05 &5.41 &5.14& 5.47&5.25 \\

  '$\top$'-shape & 5.15 & 5.40 & 5.05 & 5.41& 5.14& 5.39& 5.16 \\

  '$\bot$'-shape & 5.14 & 5.47 & 5.25& 5.48& 5.15& 5.49& 5.16 \\
\end{tabular}
\end{ruledtabular}
\end{table}

\subsection{Error analysis with typical qubit parameters}\label{SecIIIA}

\subsubsection*{Isolated CR gates}\label{SecIIIA1}

\begin{figure}[tbp]
\begin{center}
\includegraphics[keepaspectratio=true,width=\columnwidth]{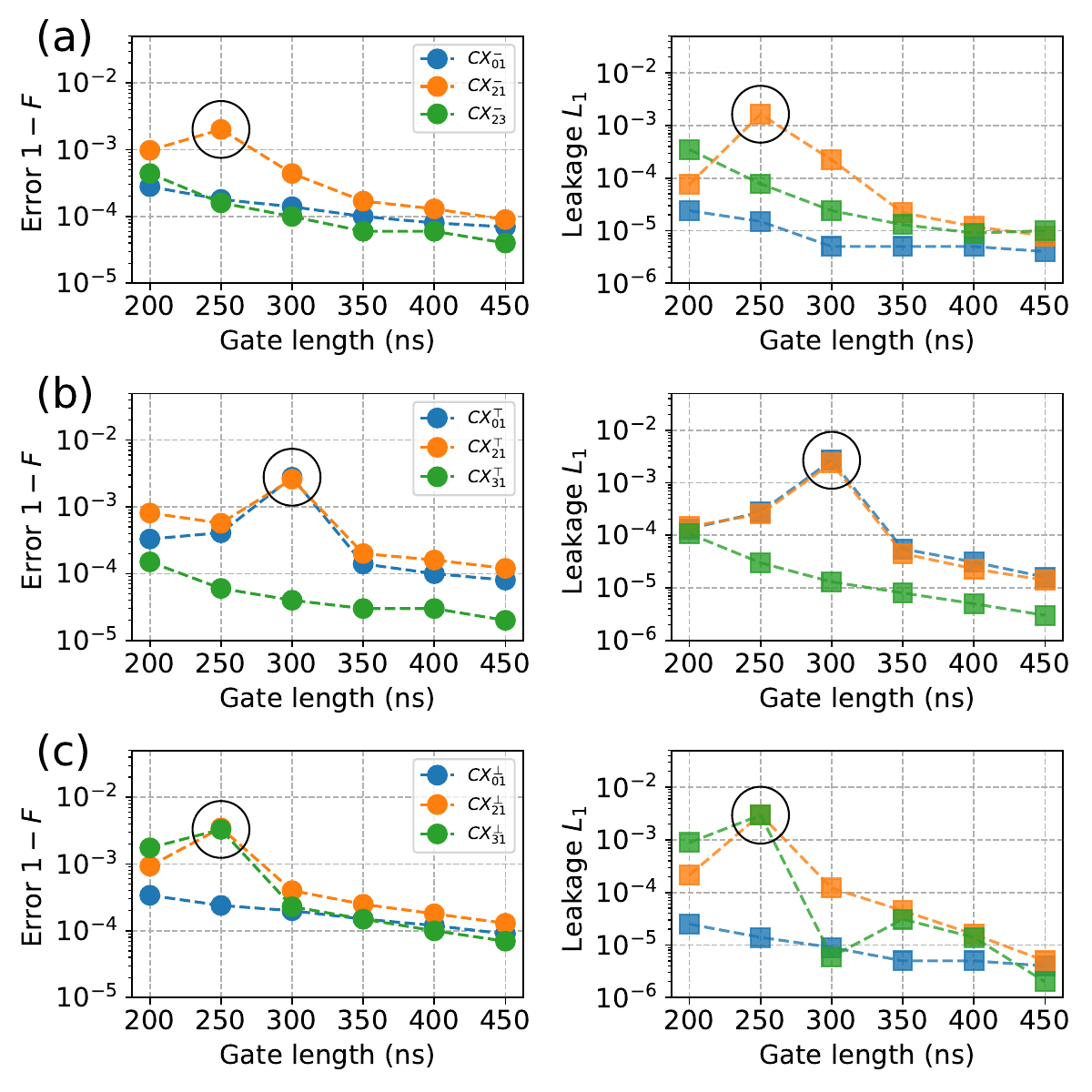}
\end{center}
\caption{The isolated gate error $1-F$ (left panel) and leakage $L_{1}$ (right panel) as
functions of the gate length. (a), (b), and (c) correspond to the four-qubit system shown in
Figs.~\ref{fig5}(a),~\ref{fig5}(b), and~\ref{fig5}(c), respectively. The used parameters are
listed in the last column of Table~\ref{tab2:System parameters}. Hereafter, for example, $CX_{21}^{-}$ denotes the CX gate
for the control qubit $Q_{2}$ and the target qubit $Q_{1}$ of the '$-$'-shape
four qubit system shown in Fig.~\ref{fig5}(a). The black circles highlight the spikes of
gate errors and leakage.}
\label{fig6}
\end{figure}

\begin{figure}[tbp]
\begin{center}
\includegraphics[keepaspectratio=true,width=\columnwidth]{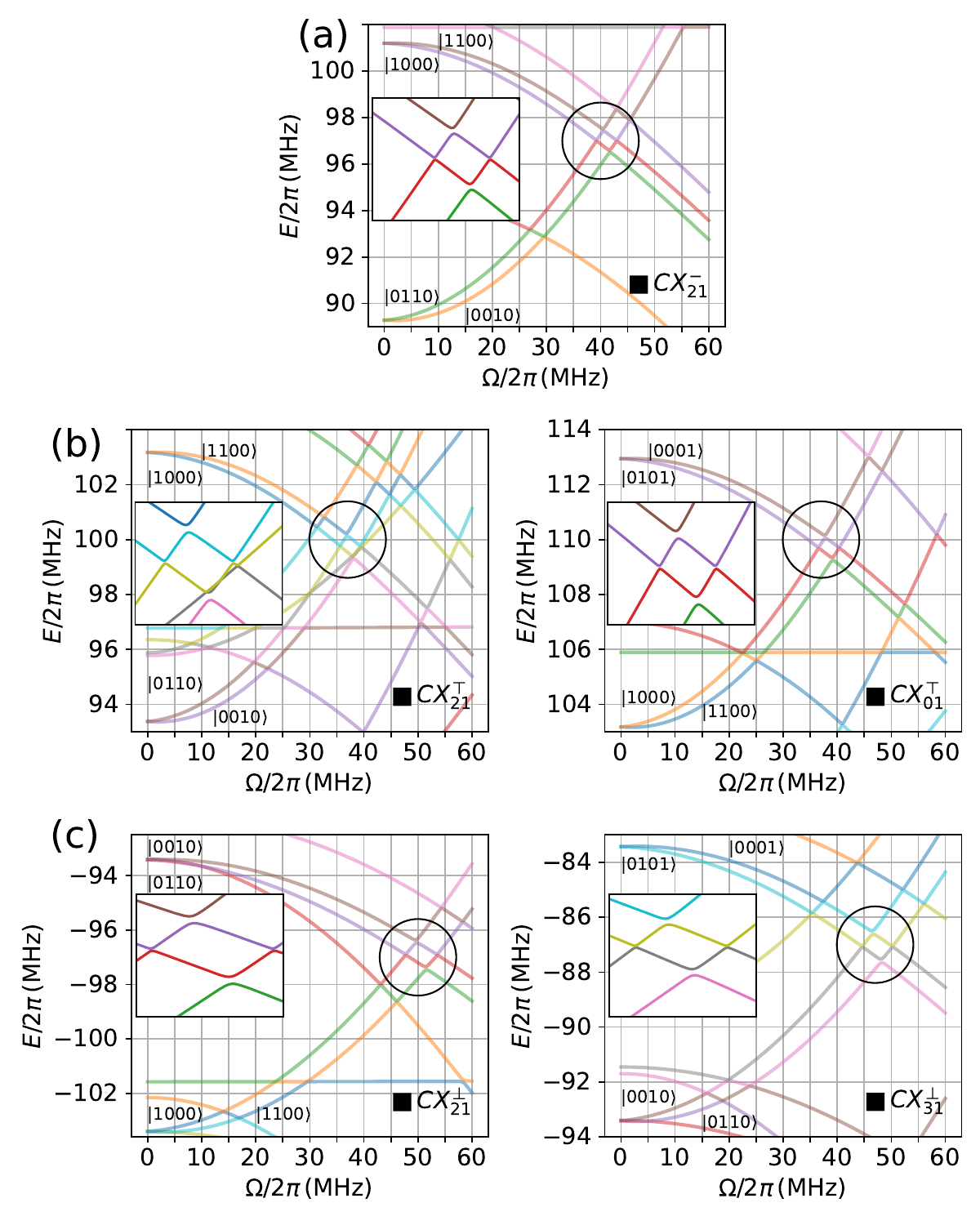}
\end{center}
\caption{Dressed spectrum of the four-qubit systems under the CR drive, showing the dressed
eigenenergies of the driven system as a function of the CR drive amplitude. Here, the
parameters used are the same as in Fig.~\ref{fig6}. (a) For the '$-$'-shape system with
the drive applied to $Q_{2}$. (b) The left and right panels denote the dressed spectrum
for the '$\top$'-shape system with the drive applied to $Q_{2}$ and $Q_{0}$, respectively.
(c) The left and right panels denote the dressed spectrum for the '$\bot$'-shape system
with the drive applied to $Q_{2}$ and $Q_{3}$, respectively. The insets enlarge the areas within the 
black circles and highlight the level anticrossings arising from next-nearest-neighbor
interactions, which lead to the spikes of gate errors and leakage shown in Fig.~\ref{fig6}.}
\label{fig7}
\end{figure}

Given the system parameters listed in Table~\ref{tab2:System parameters}, Figure~\ref{fig6} shows
both the isolated gate error $1-F$ and leakage $L_{1}$ \cite{Wood2018} as functions of the gate length for the
three four-qubit systems shown in Fig.~\ref{fig5}. Here, the ZZ coupling between near-neighbor
qubits ranges from $1\,\rm kHz$ and $13\,\rm kHz$ and is expected not to limit the following CX
gate performance. Note that hereafter the isolated gates are
characterized by assuming all the spectator qubits in their ground states \cite{Zhao2022a}, see
Appendix~\ref{C} for more details.

From the result shown in the left panels of Fig.~\ref{fig6}, as expected, increasing the gate length can generally
reduce gate errors (mainly due to the off-resonance CR drive \cite{Li2023,Malekakhlagh2022}). However, unlike isolated
two-qubit systems (see Fig.~\ref{fig4}), the gate error can show spikes (highlighted by black circles) at
specific gate lengths, which is coincident with the increase of leakage, as shown in the right
panels of Fig.~\ref{fig6}. For instance, in Fig.~\ref{fig6}(a), the gate error of $CX_{21}^{-}$ shows
a spike at the gate length of $250\,\rm ns$, coinciding with the peak of leakage. This suggests that
the spikes of the gate error dominantly arise from leakage related to spectator qubits.

Inspired by the crosstalk analysis of baseband flux-based gate operations \cite{Zhao2022a}, here we use a
similar approach based on the dressed states picture to identify the leading source of the isolated
gate errors. Note that here we focus only on the driven system with a single drive
frequency while for driven systems with multiple distinct drive frequencies (i.e., related to
simultaneous CR gates), the analysis given in Floquet theory could be devised for this
purpose \cite{Chu2004,Heya2023,Zhang2019,Gandon2022}. By moving into the rotating frame at
the CR drive frequency and within the RWA, the full system Hamiltonian, for example, describing the implementation
of $CX_{21}^{-}$ on the four-qubit system shown in Fig.~\ref{fig5}(a), can be expressed by
\begin{equation}
\begin{aligned}\label{eq21}
H=&\sum_{q}\big(\omega_{q}a_{q}^{\dagger}a_{q}
+\frac{\alpha_{q}}{2}a_{q}^{\dagger}a_{q}^{\dagger}a_{q}a_{q}\big)
+\sum_{n}\big(\omega_{r,n}a_{r,n}^{\dagger}a_{r,n}\big)
\\&+\sum_{q,n}\big[g_{r,nq}(a_{r,n}a_{q}^{\dagger}+a_{r,n}^{\dagger}a_{q})\big]+\frac{\Omega}{2}(a_{2}^{\dagger}+a_{2}),
\end{aligned}
\end{equation}
with the qubit and the resonator labeled by $q=\{0,1,2,3\}$ and $n=\{0,1,2\}$, respectively.
Here, to simplify the discussion, we neglect the cancelation drive term [see Eq.~(\ref{eq19})].

By numerical diagonalization of the Hamiltonian in Eq.~(\ref{eq21}), Figure~\ref{fig7}(a) shows the
dressed eigenenergies of the '$-$'-shape four-qubit system under the drive on $Q_{2}$ as a function of the CR
drive amplitude. One can find that near the drive amplitude of $\sim42\,\rm MHz$, there exist
level anticrossings arising from the resonance interaction between $Q_{2}$ and its next-near-neighbor 
$Q_{0}$ with the strength of $\sim 55\,\rm kHz$ (in line with the strength of
the next-nearest neighbor interaction mediated by $Q_{1}$). Meanwhile, for $CX_{21}^{-}$ with
the gate length of $250\,\rm-ns$ shown in Fig.~\ref{fig6}(a), the peak drive amplitude is
about $45\,\rm MHz$ [see also in Fig.~\ref{fig4}(a)]. These two observations indicate that during
the implementation of $250$-$\rm ns$ $CX_{21}^{-}$, the qubit system is operated
nearby the anticrossing, thus enabling the population swap between $Q_{2}$ and $Q_{0}$ and
causing the increase in leakage and gate errors. By contrast, for shorter gates (e.g., $200\,\rm ns$) or
slower gates (e.g., $300\,\rm ns$), the qubit system is operated far away from the anticrossing, allowing
the suppression of the population swap. In this situation, the gate error and leakage will be
restored to the level, which is limited mainly by the off-resonance error within the isolated two-qubit
system itself, as illustrated in Fig.~\ref{fig6}(a).

Similarly to $CX_{21}^{-}$, the spikes of gate
error and leakage in the '$\top$'-shape and '$\bot$'-shape four-qubit systems, as shown in Figs.~\ref{fig6}(b)
and~\ref{fig6}(c), also arise due to the parasitic interactions
between the control qubits and their next-near-neighbor qubits, highlighted in
Figs.~\ref{fig7}(b) and~\ref{fig7}(c). Thus, in essence, the increased gate
errors shown in Fig.~\ref{fig6} are caused by the static frequency collision resulting from the
next-nearest-neighbor interaction, as discussed in Sec.~\ref{SecIIA}.

\subsubsection*{Modeling the formation of frequency collisions}\label{SecIIIA2}

\begin{figure}[tbp]
\begin{center}
\includegraphics[keepaspectratio=true,width=\columnwidth]{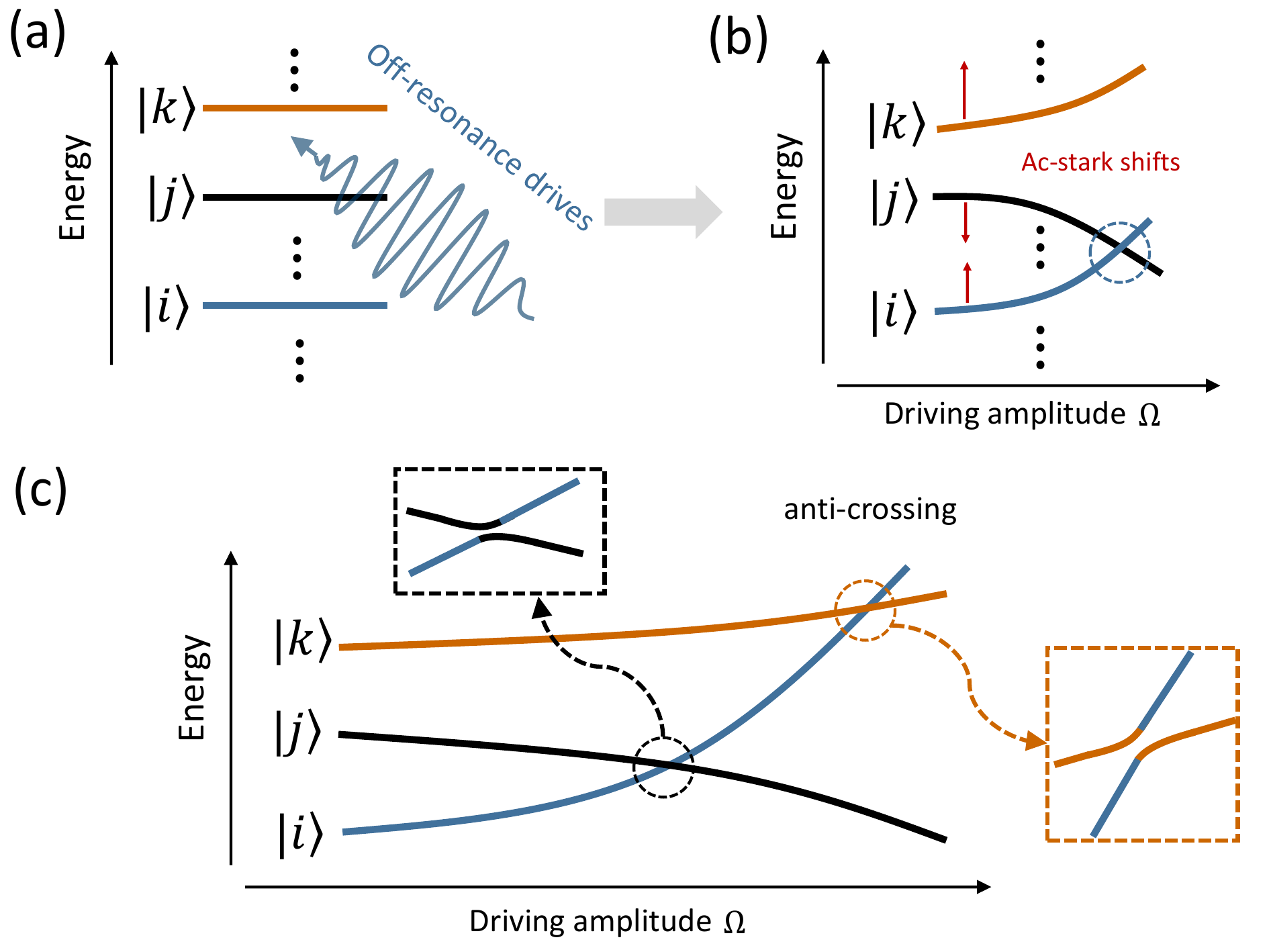}
\end{center}
\caption{Modeling the formation of frequency collisions based on the dressed state picture. (a) For a multi-level
quantum system, wherein energy levels are labeled by $|i\rangle,|j\rangle,|k\rangle...$, subjected to an off-resonance
drive, moving into the rotating frame at the drive frequency can give a dressed system. (b) The typical
dressed spectrum as a function of the drive amplitude, showing that through the ac-Stark effect, the
off-resonance drive can cause blue-shifted and red-shifted energy levels. When increasing the drive
amplitude, the shifted levels, e.g., $|j\rangle$ can sweep through the others, e.g., $|i\rangle$.
(c) Because of the interactions between levels, anticrossings (highlighted by insets) can exist at the
resonance points.}
\label{fig8}
\end{figure}

To get further insight into the underlying physics of the above collision processes, we
consider a more intuitive model based on the dressed state picture. For a multi-level
quantum system under an off-resonance drive, as shown in Fig.~\ref{fig8}(a), moving into
the rotating frame at the drive frequency gives a dressed multi-level system, whose
spectrum as a function of the drive amplitude is sketched in Fig.~\ref{fig8}(b). Through the
ac-Stark effect, the off-resonance drive can shift the levels lower or higher in frequency, which
is determined by whether or not the drive is blue-detuned or red-detuned from the transitions in
the multi-level system. As a consequence, by increasing the drive amplitude, one shifted level
can sweep through the other shifted one, as shown, for example, in Fig.~\ref{fig8}(b).
Furthermore, when the two levels are coupled together, there can exist a level anticrossing, whose
size is determined by the coupling strength, at the resonance point, as shown in the
inset of Fig.~\ref{fig8}(c).Basically, these two factors together lead to the formation of frequency collisions. In this
context, according to the primary mechanism underlying the coupling, the frequency collision
can be classified into static and dynamic, as discussed in Sec.~\ref{SecIIA}.

Similar to the above model, in the context of CR gate architecture, the off-resonance CR drive will shift
the control qubit frequency, causing accidental frequency collisions between the
control qubit and its neighbors. Specifically, we consider the collision process underlying the
results shown in Figs.~\ref{fig6} and~\ref{fig7}. As listed in Table~\ref{tab2:System parameters}, the
typical magnitude of the control-target (i.e., near-neighboring qubit pairs) detuning is $100\,\rm MHz$
while the control qubit and its next-near neighbors (i.e., another control qubit) are only
detuned by $10\,\rm MHz$. Thus, given the typical drive amplitude of $\Omega/2\pi=50\,\rm MHz$ and the detuning
magnitude of $\Delta/2\pi=100\,\rm MHz$, the ac-Stark shift can be approximated by \cite{Schneider2018}
\begin{equation}
\begin{aligned}\label{eq22}
\delta_{s}\approx \frac{\alpha_{q}\Omega^{2}}{2\Delta(\Delta+\alpha_{q})},
\end{aligned}
\end{equation}
giving rise to $16\,\rm MHz$ and $-9\,\rm MHz$ for the positive and negative detuning
conditions, respectively. Moreover, the control qubit and its next-near neighbors, e.g., $Q_{2}$ and $Q_{0}$ in
Figs.~\ref{fig6}(a) and~\ref{fig7}(a), can be coupled via the mediator, e.g., $Q_{1}$, with the typical
strength of $\sim J^{2}/\Delta$ (see also in Sec.~\ref{SecII}). Given these
considerations, the static frequency collision due to the on-resonance coupling between
neighboring control qubits can explicitly explain the presence of the spikes in gate errors and
leakage at specific drive amplitudes shown in Fig.~\ref{fig6}.

In essence, this frequency collision is caused by the always-on qubit-qubit couplings (in this sense, we
call it static frequency collision). But the condition for the occurrence of such collision depends on the CR drive
amplitude (gate length). This in turn allows one to avoid such collision issues by optimizing
drive amplitudes (gate lengths), as suggested in Fig.~\ref{fig6}.

\subsubsection*{Simultaneous gate operations}\label{SecIIIA3}

\begin{figure}[tbp]
\begin{center}
\includegraphics[keepaspectratio=true,width=\columnwidth]{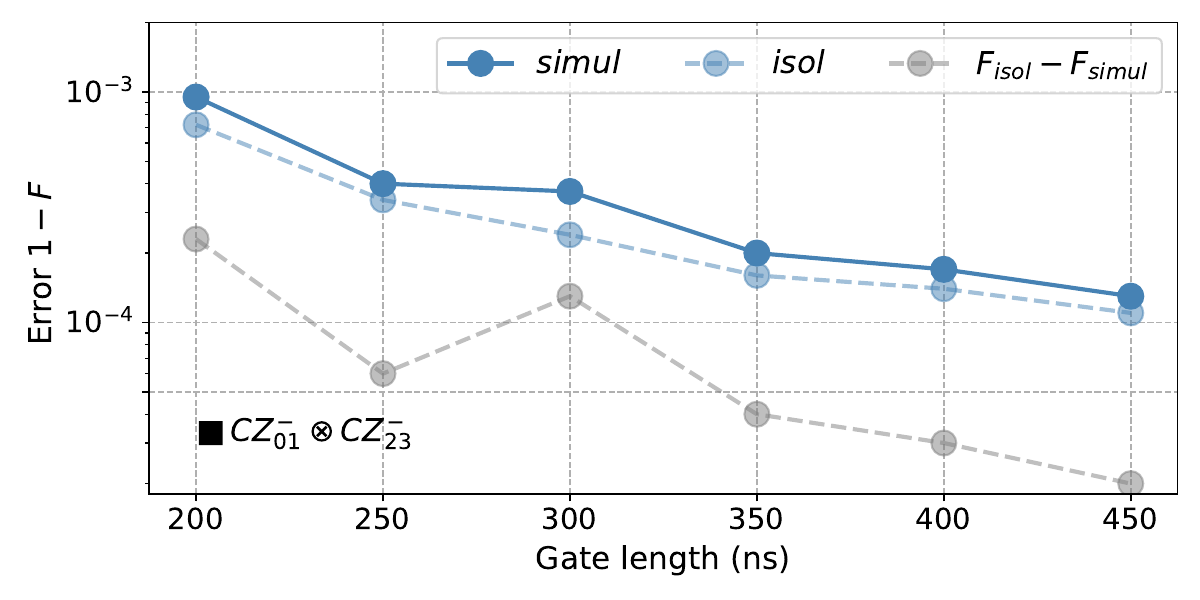}
\end{center}
\caption{Gate errors of the simultaneous CX gates ($CX_{01}^{-}\otimes CX_{23}^{-}$) in the '$-$'-shape four-qubit system
as a function of the gate length. Here, the parameters used are the same as in Fig.~\ref{fig6}(a) and the
isolated gate errors are obtained by adding the errors of the constituted isolated
gates (i.e., $CX_{01}^{-}$ and $CX_{23}^{-}$) shown in Fig.~\ref{fig6}(a).}
\label{fig9}
\end{figure}
To further examine whether there exists additional quantum crosstalk for implementing CX gates
in parallel, here we provide an analysis of the fidelity of simultaneous CX gates \cite{Zhao2022a}. With the typical
parameters listed in Table~\ref{tab2:System parameters}, Figure~\ref{fig9} shows the
performance of simultaneous gates ($CZ_{01}^{-}\otimes CZ_{23}^{-}$) in the '$-$'-shape four-qubit system
shown in Fig.~\ref{fig5}(a). Here, by adding the errors of the constituted isolated
gates $CZ_{01}^{-}$ and $CZ_{23}^{-}$, we also show the isolated
gate errors $(1-F_{\rm isol})$ and the added errors ($F_{\rm isol}-F_{\rm simul}$) when gates are implemented in parallel. Similar
to isolated gates (see Fig.~\ref{fig6}), increasing the gate length can reduce the simultaneous gate
error. Moreover, the typical added error is below $10^{-4}$, suggesting that there exists no additional
quantum crosstalk that contributes significantly to gate errors. This shows that within
the currently studied CR architecture, high-fidelity simultaneous gates may be possible, but as we will
show, this is far from the whole story. We will go back to this subject in the following section.

\subsection{Challenging from fluctuations in qubit parameters}\label{SecIIIB}

\begin{figure}[tbp]
\begin{center}
\includegraphics[keepaspectratio=true,width=\columnwidth]{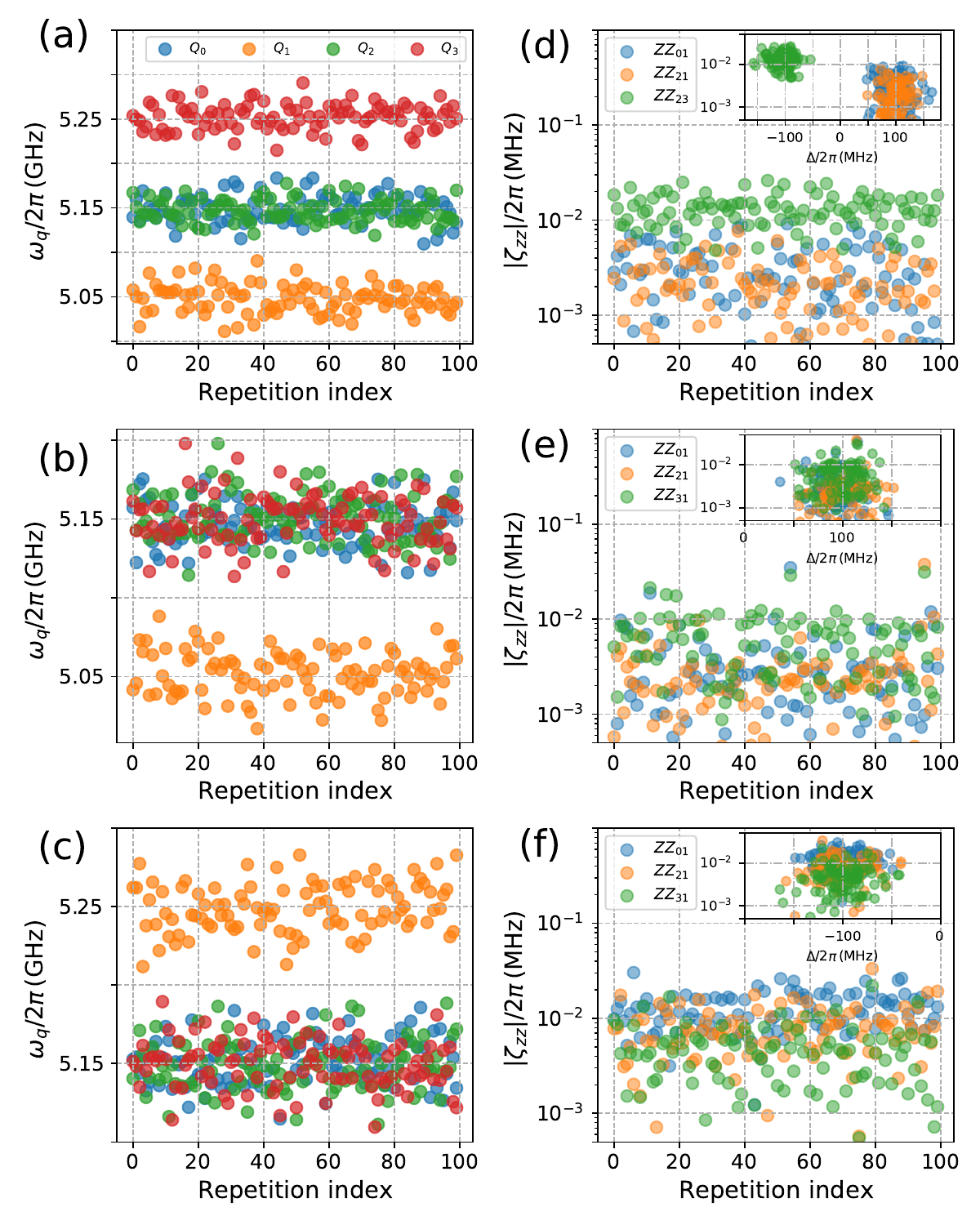}
\end{center}
\caption{Qubit frequencies and $ZZ$ couplings of $100$ repetitions. The control and target qubit frequencies
are randomly distributed around their designed values, i.e., the control frequency $\omega_{j}/2\pi=5.15\,\rm GHz$
and the target frequency $\omega_{i(k)}/2\pi=5.25(5.05)\,\rm GHz$, with the standard deviation
of $15\,\rm MHz$. (a)-(c) show the qubit frequencies of $100$ repetitions for the three four-qubit
systems shown in Figs.~\ref{fig5}(a)-(c), while the associated $ZZ$ couplings are shown in (d)-(f).
Here, other parameters are the same as in Fig.~\ref{fig6}. The insets show the distributions of the ZZ
coupling over the control-target detunings.}
\label{fig10}
\end{figure}

Below, to study the effect of frequency uncertainty on the currently studied architecture, we consider
that control and target qubit frequencies are randomly distributed around their designed values, i.e., the control
frequency of $\omega_{j}/2\pi=5.15\,\rm GHz$ and the target frequency of $\omega_{i(k)}/2\pi=5.25(5.05)\,\rm GHz$, with
the standard deviation of $\sigma_{f}=15\,\rm MHz$ \cite{Hertzberg2021}. With $100$ distinct repetitions, we provide analysis
of both the $ZZ$ suppression and the performance of $CX$ gates (including both isolated gates and simultaneous gates) for
the four-qubit systems shown in Fig.~\ref{fig5}. Accordingly, Figures~\ref{fig10}(a)-(c) show the qubit frequencies of $100$ repetitions
for the four-qubit systems shown in Figs.~\ref{fig5}(a)-(c).

\subsubsection*{$ZZ$ suppression}\label{SecIIIB1}

As shown in Figs.~\ref{fig3}, for isolated two-qubit systems with the desired setting
in qubit frequencies, the $ZZ$ coupling can be suppressed below $10\,\rm kHz$.
However, as we will show, the deviation from the target setting in multi-qubit systems will make the $ZZ$
suppression more complicated.

Figures~\ref{fig10}(d)-(f) show the ZZ couplings of $100$ repetitions and the maximum (median) values
of the $ZZ$ coupling are $26\,(3)\,\rm kHz$, $38\,(3)\,\rm kHz$, and $33\,(7)\,\rm kHz$ for the
four-qubit systems shown in Figs.~\ref{fig5}(a)-(c), respectively. Moreover, it is also shown that
the 'exceptional' points with large $ZZ$ couplings almost appear in pairs, as shown in
Fig.~\ref{fig10}(e). To check their dependence on the qubit frequencies, the inset
also shows the distributions of the ZZ coupling over the control-target detunings. We
find that these 'exceptional' points arise mainly from static frequency collisions due
to interactions between next-nearest-neighboring control qubits.

To be more specific, similar to the frequency collision of (S1), when the control qubit is on-resonance
with its neighboring control qubit, the strong state hybridization due to the next-nearest-neighboring couplings
can degrade qubit addressability and result in prominent quantum crosstalk.
This is reasonable as the frequencies of all the control qubits are distributed around the same
value, making them more prone to this frequency-collision issue. For instance, in Fig.~\ref{fig10}(e), the
maximum $ZZ$ coupling of $38\,\rm kHz$ is caused by the nearly on-resonance coupling of neighboring
control qubits $Q_{2}$ and $Q_{3}$. By excluding these 'exceptional' points
arising from these static collisions, we find that the typical $ZZ$ coupling could
be below $20\,\rm kHz$. Similar to the deterioration of $ZZ$ suppression, as we will
show, such collisions can also limit the CX gate performance and cause
spikes of gate errors appearing in pairs.

\subsubsection*{Isolated CX gates}\label{SecIIIB2}

\begin{figure*}[tbp]
\begin{center}
\includegraphics[width=16cm,height=10cm]{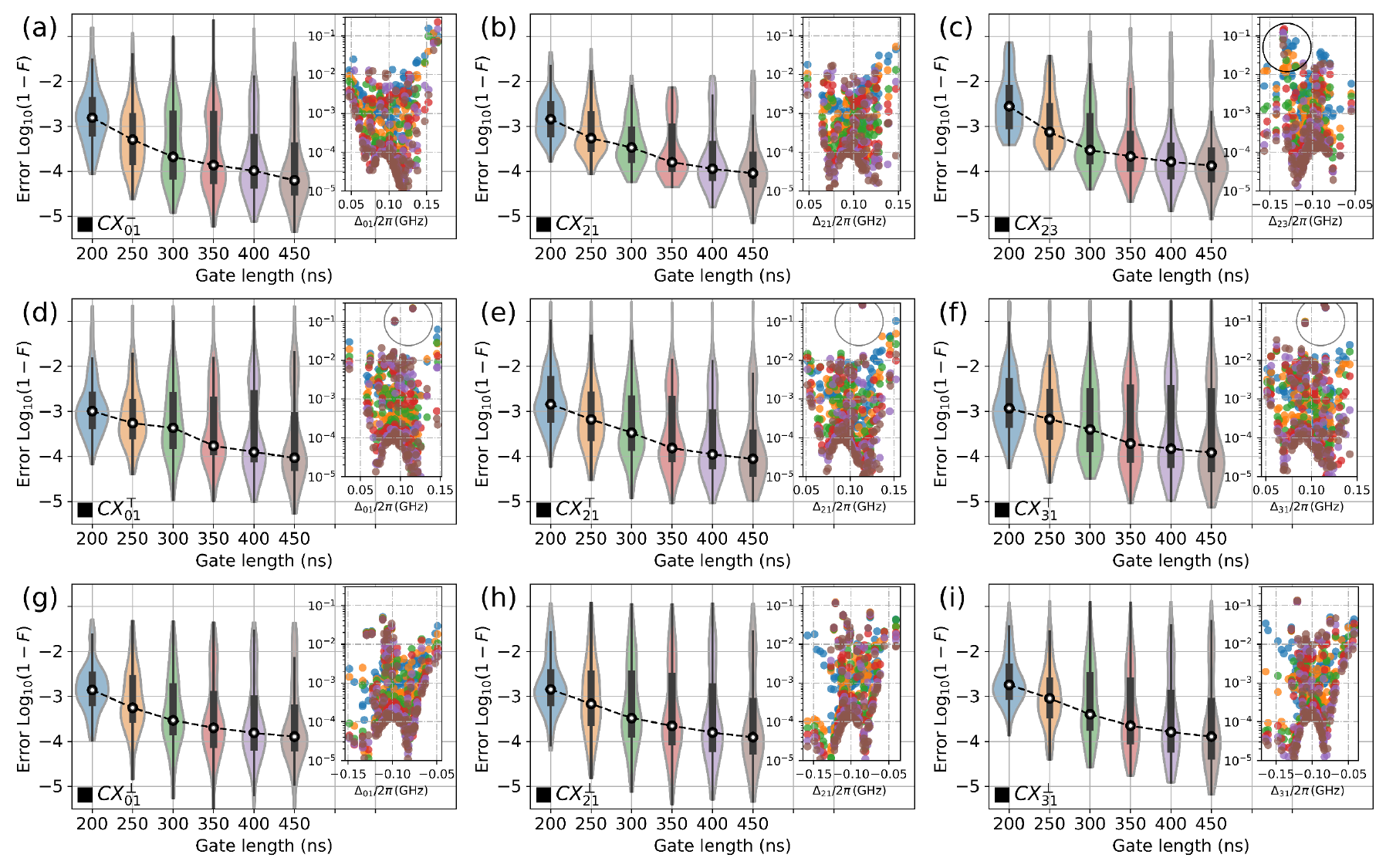}
\end{center}
\caption{Isolated CX gate errors versus gate lengths of $100$ repetitions. Here, the used parameters are the same as
in Fig.~\ref{fig10}. The violin plots in (a-c), (d-f), and (g-i) show the distributions of isolated
gate errors versus gate lengths for the four-qubit systems shown in Figs.~\ref{fig5}(a),~\ref{fig5}(b), and~\ref{fig5}(c),
respectively. The dashed lines denote the median gate error as a function of the gate length
and the insets give the error distributions over the control-target detunings. The black (grey) circles
highlight the spikes of gate errors arising from frequency-collision issues.}
\label{fig11}
\end{figure*}

The violin plots in Fig.~\ref{fig11} show the distributions of isolated CX gate errors with varying
gate lengths. Similar to isolated two-qubit systems (see Fig.~\ref{fig4}), here the
median gate errors (indicated by dashed lines) are reduced by increasing the gate length
and the typical gate error ranges from $10^{-4}$ to $10^{-3}$. Moreover, as shown in the inset
of Fig.~\ref{fig11}, when increasing the gate lengths, almost all qubits show decreasing trends in
gate errors, suggesting that the gate error mainly comes from the CR drive. Besides, three main types
of features deserve to be discussed in more detail.

(i) As shown, for example, in Fig.~\ref{fig11}(a), under the positive detuning condition, large gate errors
can be found near the detunings of $50\,\rm MHz$ and $150\,\rm MHz$. Following the discussion in Sec.~\ref{SecIIA}, here the
gate error is mainly dominated by the off-resonance transitions, i.e., the transition $|0\rangle\leftrightarrow|1\rangle$
and the two-photon transition $|0\rangle\leftrightarrow|2\rangle$ of the control qubit, which
correspond to the frequency collisions of (S1) and (D1). As expected, prolonging the
gate length (i.e., decreasing the drive amplitude) or increasing ramp time (i.e., mitigating non-adiabatic
transitions in the dressed picture, as discussed in Sec.~\ref{SecIIIA}), these transitions can be suppressed.
Furthermore, as already discussed in previous works \cite{Li2023,Malekakhlagh2022}, we expect that off-resonance errors shown
in Fig.~\ref{fig11} can be largely suppressed by optimizing the control pulse shape in Eq.~(\ref{eq20}).

In addition, as shown in Figs.~\ref{fig11}(g)-(i), under the negative detuning conditions, large gate errors
can also be found near the detunings of $50\,\rm MHz$. Similar to that within positive detuning conditions,
this arises from the transition $|0\rangle\leftrightarrow|1\rangle$ related to the frequency collision (S1).
Meanwhile, in contrast, as here the control-target detuning is negative, the two-photon
transition $|0\rangle\leftrightarrow|2\rangle$ is significantly suppressed by the large detuning
of $\sim 265\,\rm MHz$. Thus, with the negative detuning condition, the frequency-collision
issue of (D1) can be safely omitted for ensuing high-fidelity gates.

\begin{figure}[tbp]
\begin{center}
\includegraphics[keepaspectratio=true,width=\columnwidth]{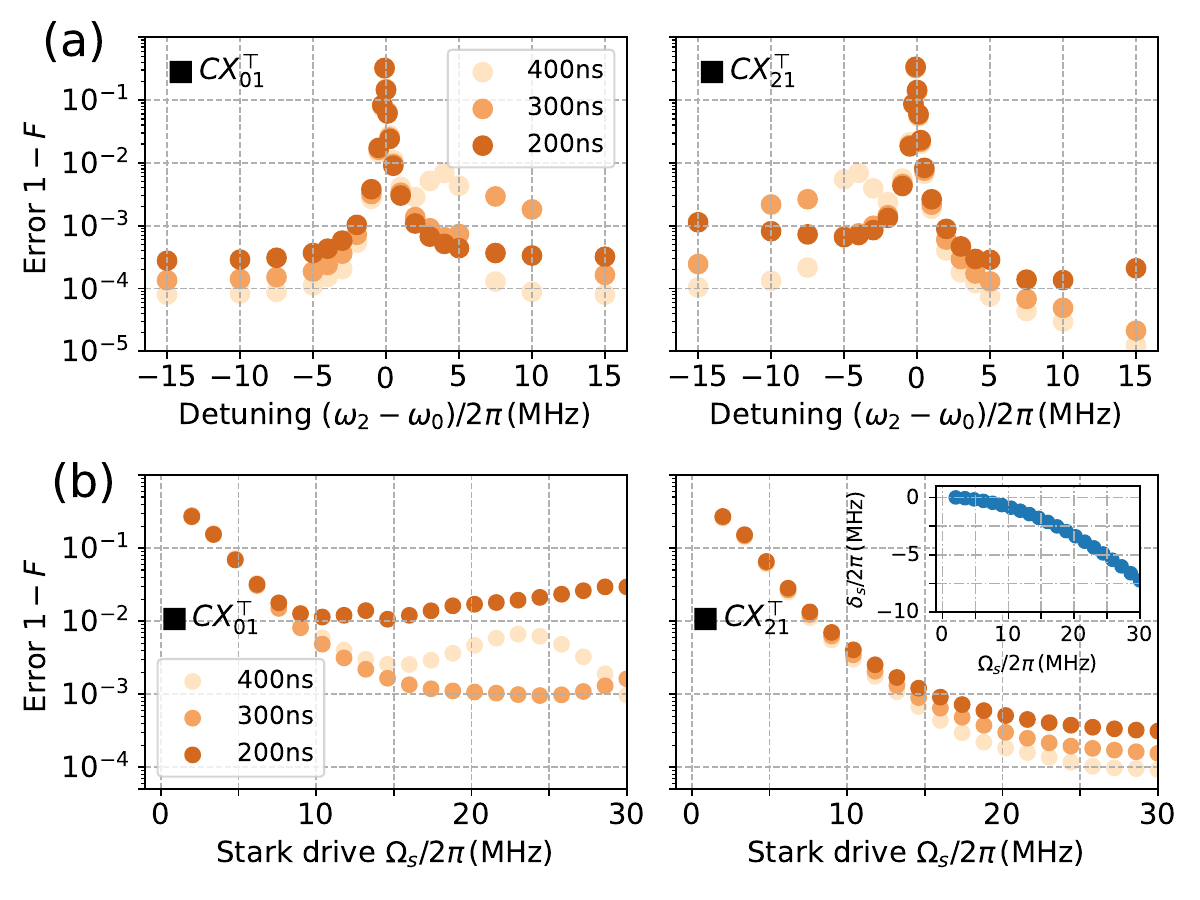}
\end{center}
\caption{(a) CX gate errors versus the detuning between neighboring control qubits ($Q_{0}$ and $Q_{2}$) with different gate
lengths for a three-qubit subsystem of the '$\top$'-shape system. Here, the qubit frequencies of $Q_{0}$ and $Q_{1}$
are $5.15\,\rm GHz$ and $5.05\,\rm GHz$, respectively, and the other parameters are the same as
in Figs.~\ref{fig11}(d) and~\ref{fig11}(e). (b) CX gate errors with the presence of an always-on off-resonance
drive applied to $Q_{0}$. Here, the frequency of $Q_{2}$ is $5.15\,\rm GHz$ and the Stark drive frequency
is $5.20\,\rm GHz$. The inset shows the ac-Stark shift $\delta_{s}$ versus the drive amplitude $\Omega_{s}$.}
\label{fig12}
\end{figure}

(ii) Similar to the prominent $ZZ$ crosstalk, indeed, the static collision arising from on-resonance coupling
of neighboring control qubits can account for the spikes of gate errors appearing in pairs, as highlighted, for
example, by the grey circles in Figs.~\ref{fig11}(d)-(f). More importantly, different from the dynamic collisions
and the static collision discussed in Figs.~\ref{fig6} and~\ref{fig7}, the current collision
is almost independent of the CR drive, explaining why the resulting gate errors are not reduced by increasing the gate length, as
shown in Figs.~\ref{fig11}(d)-(f). More specifically, Figure~\ref{fig12}(a) shows the CX gate errors versus the detuning
between neighboring control qubits with different gate lengths for a subsystem (consisting of $Q_{0}$, $Q_{1}$, and $Q_{2}$)
of the '$\top$'-shape qubit system. One can find that when the two control qubits are on-resonance, the spikes of
gate errors appear in pairs (see the left and right panels) and peak values are almost independent of
the gate length. Additionally, it is also shown that since the typical coupling strength between neighboring
control qubits is of the order of $50\,\rm kHz$ (see in Sec.~\ref{SecIIIA}), a small detuning from the frequency
collision, i.e., $\sim0.5\,(2.0)\,\rm MHz$, is adequate to ensure gate error of $\sim 0.01$ ($0.001$).

Besides, as one might expect, the on-resonance coupling of control qubits can lead to worsening qubit addressability
and will inevitably complicate almost all qubit operations (such as qubit readout), not only restricted to gate
operations. It is in this sense that we argue that this type of static frequency collision deserves to be addressed
with a high priority. In the current CR architecture, to avoid such static collision, one possible approach
is by adding an additional always-on off-resonant drive to selectively shift qubits \cite{Zhao2022c,Wei2022}.
As shown in Fig.~\ref{fig12}(b), by applying a stark drive to $Q_{0}$ with the detuning of $50\,\rm MHz$, ensuring
gate errors of $0.001$ only requires an amplitude of $\sim 15\,\rm MHz$. As shown in the inset
of Fig.~\ref{fig12}(b), this drive induces a frequency shift of $\sim 1.9\,\rm MHz$, agreeing well with
the estimated collision bound [see Fig.~\ref{fig12}(a)]. Additionally, as already studied in the previous
work \cite{Zhao2022c}, the Stark drive itself could also introduce additional error sources, this can explain
why generally the performance of $CX_{01}^{\top}$ is worse than that of $CX_{21}^{\top}$, as shown in Fig.~\ref{fig12}(b). 
Moreover, with the mitigation of the frequency-collision issue, the
qubit addressability could be improved and high-fidelity readout could also be possible, even under the
always-on off-resonance drive \cite{Zhao2023,Chen2022}. Similarly, as shown in Figs.~\ref{fig11}(g)-(i), such
static collision issues also exist for the '$\bot$'-shape system, and, in Appendix~\ref{D}, we also
show the results for addressing this issue based on the ac-Stark effect.

(iii) Compared to other existing CR architectures listed in Table~\ref{tab1:CR architectures}, the most peculiar
feature of the currently studied architecture is that the magnitude of the qubit-resonator detuning is
comparable to that of the control-target detuning, as well as that of the qubit anharmonicity. This can
raise a crucial question, whether this setting will introduce an additional leakage channel or error
source. To answer that question, similar to the discussion of frequency collision for qubits given
in Sec.~\ref{SecIIA}, it should be necessary to analyze the frequency-collision issue associated
with both the qubits and the resonators, which could potentially cause leakage and gate errors.
Since the typical detuning between the qubit and its resonator coupler is larger than $250\,\rm MHz$, frequency
collisions similar to that given in Sec.~\ref{SecIIA}, i.e., (S1-S2) and (D1-D4), can be ignored for
the present purpose. Moreover, as it is generally assumed that all the resonator couplers
are initialized in their ground states, the frequency collision should be enabled by
the CR drive-induced multiphoton process, thus acting as a dynamic one.

\begin{figure}[tbp]
\begin{center}
\includegraphics[keepaspectratio=true,width=\columnwidth]{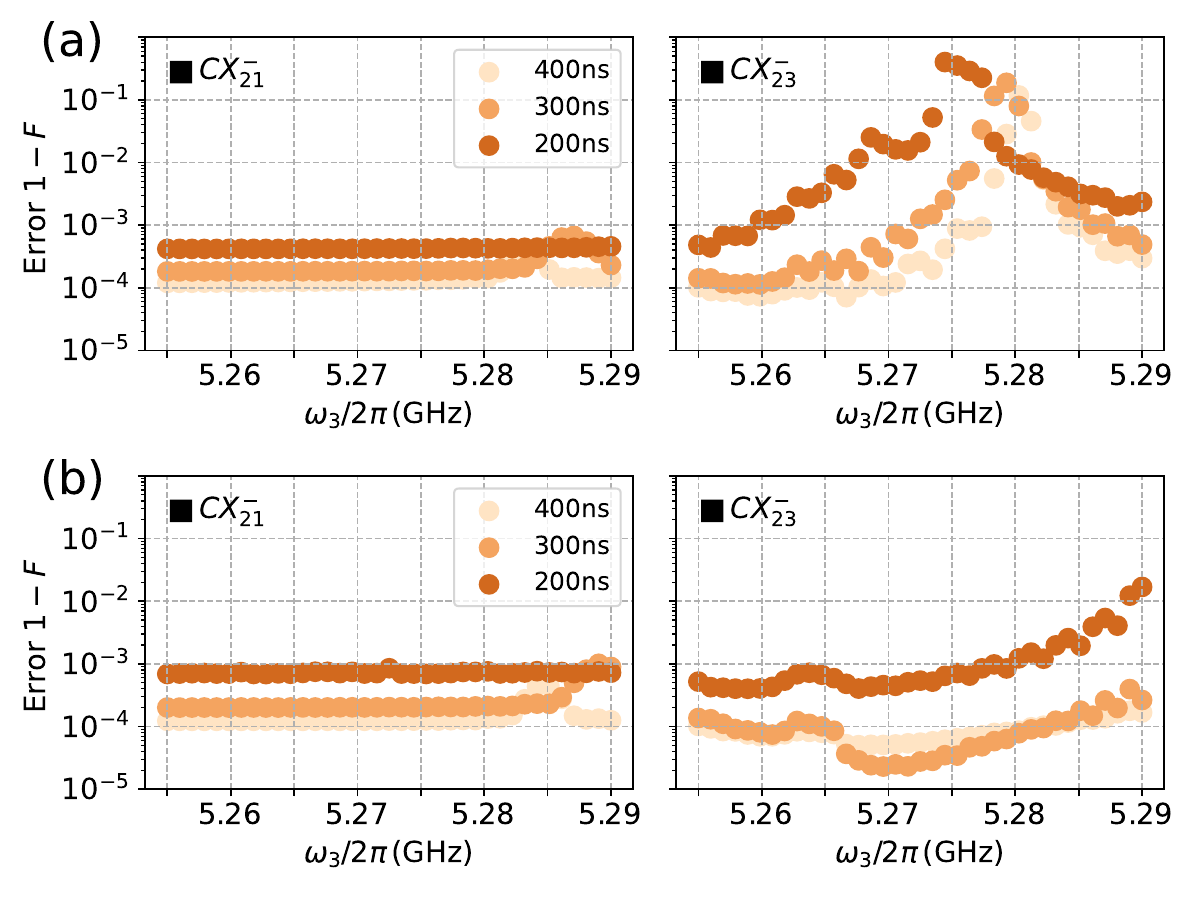}
\end{center}
\caption{(a) CX gate errors as a function of the frequency of $Q_{3}$ with different gate
lengths for a three-qubit subsystem of the '$-$'-shape system. Here, the qubit frequencies of $Q_{1}$ and $Q_{2}$
are $5.05\,\rm GHz$ and $5.15\,\rm GHz$, respectively, and the other parameters are the same as
in Fig.~\ref{fig11}(c). (b) CX gate errors with increased frequency of the
lightweight resonator. Here, the frequency of resonator $R_{1}$ is $5.45\,\rm GHz$.}
\label{fig13}
\end{figure}

Accordingly, considering the present frequency setting (i.e., $\omega_{i}>\omega_{j}>\omega_{k}$) and
the typical parameters listed in Table~\ref{tab2:System parameters}, we can expect
that the leading frequency collision should satisfy the following relation $D_{r}$ (see Appendix~\ref{E} for details)
\begin{equation}
\begin{aligned}\label{eq23}
\omega_{r}+\omega_{j}=2\omega_{i}.
\end{aligned}
\end{equation}
This describes the two-photon transition $|0000\rangle\leftrightarrow|1100\rangle$ in the
coupled qubit-resonator system, wherein the state is denoted by $|R_{1},Q_{j},R_{2},Q_{i}\rangle$, and the
transition rate can be approximated by (applying second-order perturbation theory) \cite{Tripathi2019,Nesterov2021,Poletto2012}
\begin{equation}
\begin{aligned}\label{eq24}
V_{00\leftrightarrow11}\simeq\frac{2g_{rq}\Omega^{2}\alpha_{q}}{(\alpha_{q}+\Delta_{q})\Delta_{q}^{2}}.
\end{aligned}
\end{equation}
Similarly, the two-photon transition $|0100\rangle\leftrightarrow|1200\rangle$ with
the condition of $\omega_{r}+\omega_{j}+\alpha_{j}=2\omega_{i}$ can occur, but the transition is
heavily suppressed by the large qubit-resonator detuning $\Delta_{rq}$ and the negative anharmonicity
of transoms in our setting \cite{Zhao2022a} (see Appendix~\ref{E} for
details).

Indeed, the dynamic frequency collision $D_{r}$ causes the spikes of gate errors highlighted by
the black circle in the inset of Fig.~\ref{fig11}(c). To provide a more detailed analysis of the
collision issue, Fig.~\ref{fig13}(a) shows the gate errors as a function of $Q_{3}$ for a
subsystem (consisting of $Q_{1}$, $Q_{2}$, and $Q_{3}$) of the '$-$'-shape qubit system. As
expected [see Eq.~(\ref{eq23})], the presence of this dynamic collision does not
affect the performance of $CX_{21}^{-}$ while it can significantly degrade the $CX_{23}^{-}$ gate
performance. Moreover, it also illustrates the typical features of dynamic frequency collisions, i.e., both
the collision condition and the induced gate error (transition rate) show a dependence on the CR drive
amplitude (gate length). This suggests that, to avoid the detrimental impact of such collision, one can optimize
the gate length, similar to the case discussed in Fig.~\ref{fig6}. Besides, according
to Eq.~(\ref{eq23}), increasing the resonator frequency can be an alternative approach for
this purpose, as shown in Fig.~\ref{fig13}(b), which shows the CX gate errors with a larger
resonator frequency. Additionally, in Appendix~\ref{D}, we also show that such collisions can also
be mitigated through the ac-Stark shift, similar to the one illustrated in Fig.~\ref{fig12}(b).

\subsubsection*{Simultaneous CX gates}\label{SecIIIB2}

\begin{figure}[tbp]
\begin{center}
\includegraphics[keepaspectratio=true,width=\columnwidth]{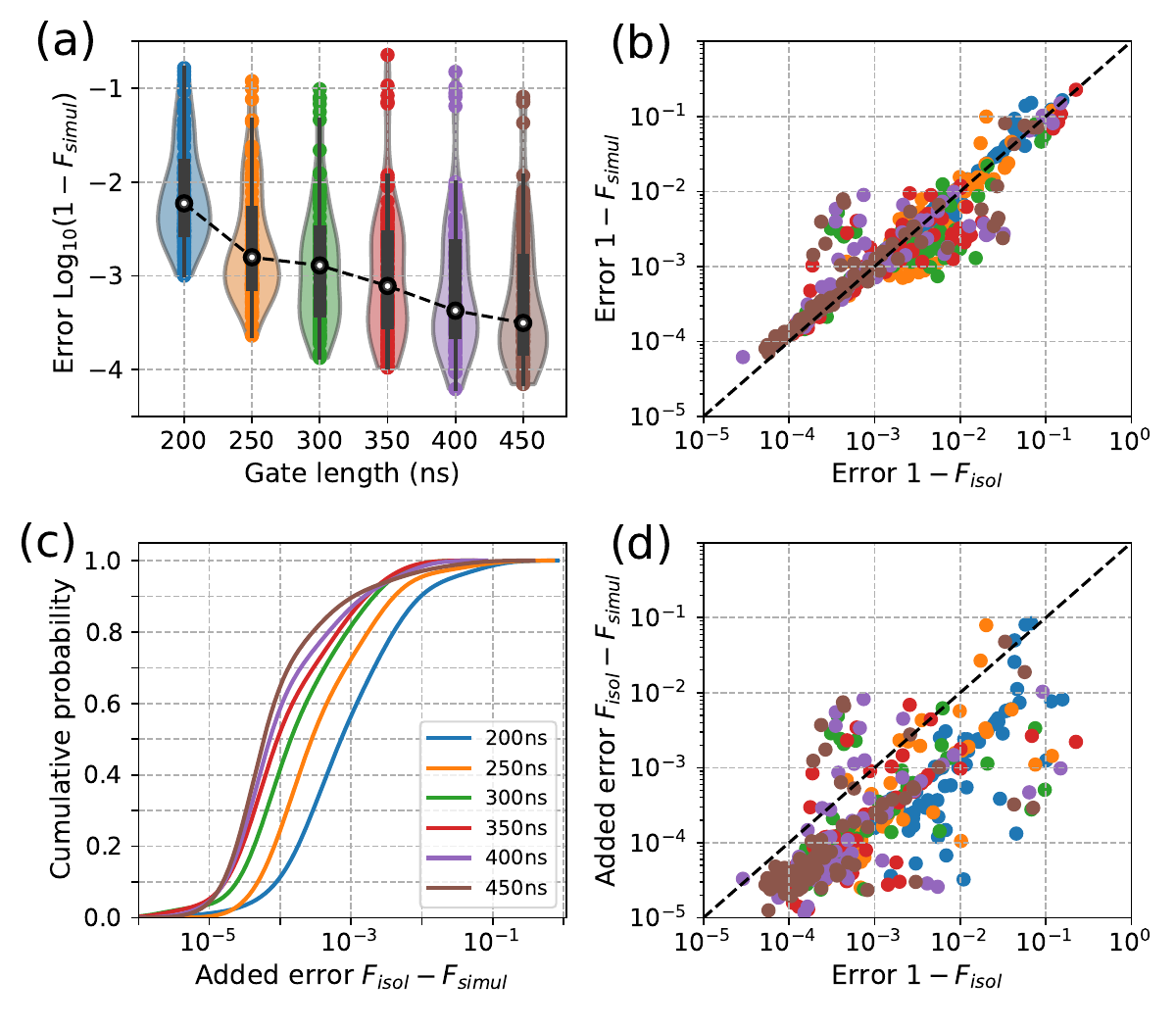}
\end{center}
\caption{ Gate errors of the simultaneous CX gates ($CX_{01}^{-}\otimes CX_{23}^{-}$) in the '$-$'-shape four-qubit system
versus gate lengths of 100 repetitions. Here, the parameters used are the same as in Figs.~\ref{fig11}(a)
and~\ref{fig11}(c). (a) Violin plot of the distributions of simultaneous gate errors ($1-F_{\rm simul}$) versus
gate lengths. The dashed lines denote the median gate error as a function of the gate length. (b) Scatter plot
of the distributions of simultaneous gate errors over the isolated errors. Here, the isolated error is obtained
by adding the errors of the constituted isolated gates in Figs.~\ref{fig11}(a) and~\ref{fig11}(c). (c) Cumulative
distribution of the added gate errors $F_{\rm isol}-F_{\rm simul}$ (excluding negative values, see text). (d) Scatter plot
of the distributions of the added gate errors over the isolated errors.}
\label{fig14}
\end{figure}

Here, we turn to study CX gates implemented in parallel, especially focusing on
simultaneous CX gates in the '$-$'-shape four-qubit system shown in
Fig.~\ref{fig5}(a). In Fig.~\ref{fig14}(a), the violin plot shows the distributions of
simultaneous gate errors $1-F_{\rm simul}$ versus the gate length. Similar to that shown in
Fig.~\ref{fig9}, the median gate errors (indicated by dashed lines) decrease when
increasing the gate length, in line with the expectation that the dominant gate
error comes from the CR drive. Moreover, the simultaneous gate errors show a clear linear
dependence on the isolated gate errors, as shown in Fig.~\ref{fig14}(b).

Besides, a close look at Fig.~\ref{fig14}(b) also shows us that while the simultaneous gate performance is generally
worse than that of the isolated one, in some cases (especially for isolated gates with the
fidelities limited by leakage), the simultaneous gate can show better
performance. As already noted in the previous work \cite{Zhao2022a}, here the isolated gates are
characterized by assuming spectator qubits in ground states. Thus leakage occur during
isolated gate operations (such as, see Fig.~\ref{fig6}, when qubits are on resonance with spectator
qubits) will contribute to the control error of simultaneous gates. In this
sense, the presence of such a situation is partly due to our ill-defined metric and thus the added
error is only indicative. Excluding such cases, Figure~\ref{fig14}(c) shows the cumulative
distribution of the added gate errors $F_{\rm isol}-F_{\rm simul}$. It is shown that the median added
error is below $10^{-3}$ and can further reduce by increasing the gate length. For instance,
with a gate length of $350\,\rm ns$, the median error can be suppressed
below $10^{-4}$.

To further evaluate the effect of quantum crosstalk residing in the currently studied architecture,
in Fig.~\ref{fig14}(d), we also show the distributions of added errors over the isolated
gate errors. One can find that in almost all the cases, the added errors are smaller than that of
the associated isolated gates. This suggests that in the currently studied architecture, there exists
no additional quantum crosstalk that contributes significantly to gate errors for $CX$ gates
implemented in parallel.

\section{Discussion}\label{SecIV}

In this section, we briefly summarize the main frequency-collision issue in the currently studied
CR architecture and give potential approaches to address this issue. Then, we provide a discussion
of parameter sensitivity in the current architecture.

\subsubsection*{Revisiting the frequency collision issue}\label{SecIVA}

As discussed in Sec.~\ref{SecIIA} and Sec.~\ref{SecIII}, according to the underlying mechanism, the
frequency collision can be classified into two main types, static and dynamic.
Accordingly, near frequency-collision regions, the two types of collisions can be described by the following
effective two-level Hamiltonian given in the dressed state picture (see Fig.~\ref{fig8})
\begin{equation}
\begin{aligned}\label{eq25}
H_{sc}=\frac{\delta_{s0}+\delta_s(\Omega)}{2}Z+\frac{\lambda_{s}}{2}X,
\end{aligned}
\end{equation}
\begin{equation}
\begin{aligned}\label{eq26}
H_{dc}=\frac{\delta_{d0}+\delta_d(\Omega)}{2}Z+\frac{\lambda_{d}(\Omega)}{2}X,
\end{aligned}
\end{equation}
where $H_{sc}$ and $H_{dc}$ are for describing the static and the dynamic collisions, respectively.
Here, $\delta_{s0}$ ($\delta_{d0}$) denotes the two-level detuning without the CR drive while
$\delta_s(\Omega)$ ($\delta_d(\Omega)$) is the ac-Stark shift, generally displaying a quadratic
dependence on the drive amplitude $\Omega$ [see Eq.~(\ref{eq22})]. $\lambda_{s}$ denotes the coupling
enabling the static collisions and is almost independent of the drive amplitude, while
$\lambda_{d}(\Omega)$ is the coupling for the dynamic collision and
strongly depends on the CR drive amplitude. Below, based on Eqs.~(\ref{eq25})
and~(\ref{eq26}), we give more detailed discussions on the frequency collision:

(i) The static frequency collision. As in Eq.~(\ref{eq25}), the static collision and the associated
frequency bound mainly depend on the static couplings $\lambda_{s}$, such as couplings between 
near-neighbors or beyond. To mitigate such collisions, a finite detuning of $\delta_{s0}$ should
be retained with the highest priority. Otherwise, this will worsen qubit addressability and significantly
complicate almost all of the qubit operations. On the other hand, even with the predefined
detuning $\delta_{s0}$, the CR drive can shift qubits, giving rise to the additional
detuning $\delta_s(\Omega)$. This means that the on-resonance condition can be achieved
at specific drive amplitudes, as shown in Figs.~\ref{fig6} and~\ref{fig7}. Thus, to further avoid such situations, the drive
amplitude (gate length) should be optimized, keeping away from the on-resonance
point (see Figs.~\ref{fig6} and~\ref{fig7}). In the currently studied CR architecture, the
collision issue for neighboring qubits can be largely mitigated by using the three-frequency
allocation scheme on the heavy-hexagonal qubit lattice, as illustrated in Figs.~\ref{fig1}(a) and~\ref{fig1}(e).
However, given the three-frequency pattern, neighboring control qubits are
prone to such collision issue. Hopefully, the typical coupling is at the
level of sub-MHz, thus, according to Eq.~(\ref{eq25}), a small detuning is adequate
for addressing such issues [see Fig.~\ref{fig12}(a)]. Meanwhile, even without
the predefined detuning $\delta_{s0}$, adding a weak off-resonant drive to selectively
shift qubits can mitigate this issue, as demonstrated in Fig.~\ref{fig12}(b).

(ii) The dynamic frequency collision. From Eq.~(\ref{eq26}), both the detuning and the
coupling depend on the CR drive amplitude \cite{Zhang2019}. In essence, this issue is generally
manifested as drive-induced single- or multi-photon transitions, which are absent when the
drive is turned off, and the main types are summarized in Sec.~\ref{SecIIA}. Accordingly, both
the resulting gate error and the occurrence condition can show a clear dependence on the
drive amplitude, as shown, for example, in Fig.~\ref{fig13}(a). Thus, generally, to mitigate
the issue, one can keep a predefined detuning $\delta_{d0}$ and decrease the drive amplitude.
On the other hand, besides improving the accuracy in setting $\delta_{d0}$, one
might prefer to optimize the control pulse for ensuring fast-speed
gates \cite{Li2023,Malekakhlagh2022}. In the dressed state picture, this corresponds to
suppressing the non-adiabatic transitions which are dominated by the collisions with large
couplings. Note that similar to baseband flux-control gate operations \cite{Zhao2022a}, in the
long term, one should balance gate errors due to frequency collisions with small
couplings (sub-MHz, favoring fast gates) and with large couplings (few MHz, preferring slow gates).

In the present architecture based on heavy-hexagonal layout and the allocation scheme shown
in Figs.~\ref{fig1}(a) and~\ref{fig1}(e), the dynamics-collision issue for neighboring qubits
can be largely mitigated by assuming the state-of-the-art accuracy in setting qubit
frequencies \cite{Hertzberg2021}, as shown in Fig.~\ref{fig10}. But indeed, new frequency collisions can be introduced here, as
indicated by Eq.~(\ref{eq23}), due to the small qubit-resonator detuning (note that as shown in
Fig.~\ref{fig2}(c), within the architecture with the multipath coupler, this issue should
be heavily suppressed). In Fig.~\ref{fig13}(b), we show that increasing the resonator
frequency, i.e., increasing $\delta_{d0}$ in Eq.~{\ref{eq26}}, can address this issue while it does
not seriously affect the gate speed [see also Fig.~\ref{fig4}(a)]. Additionally, as illustrated
in Appendix~\ref{D}, this issue can also be mitigated by adding off-resonance drives. Note that,
unlike the static collision, here the dynamics collision is activated by the CR drive. Thus, in
principle, the off-resonance drive could only be applied during gate operations \cite{Zhao2022c,Wang2023},
rather than being applied in an always-on manner. In this scenario, to ensure high-fidelity
state maps between the computational basis and microwave-dressed basis, the Stark drive
should be slowly ramped up or down \cite{Zhao2022c}.

\subsubsection*{Parameter sensitivity}\label{SecIVB}

As illustrated in Sec.~\ref{SecII} and Sec.~\ref{SecIII}, in the currently studied architecture,
the realization of fast-speed CR agates with $ZZ$ suppression depends strongly on the accuracy
in setting qubit parameters. For instance, assuming the control-target detuning of $100\,\rm MHz$ and the
$ZZ$ coupling suppressed below $10\,\rm kHz$, the typical usable range of qubit-resonator detuning for ensuring
fast-speed gates (here we consider that, assuming a CR drive amplitude of $50\,\rm MHz$, the
target $ZX$ rate is $1\,\rm MHz$), is only about $100\,\rm MHz$, as shown
in Figs.~\ref{fig2}(d) and~\ref{fig3}. Considering the fabrication uncertainty, within the current
architecture, ensuring the successful realization of fast gates with $ZZ$ suppression should
be a non-trivial task.

Generally, there are two kinds of solutions for this issue: (i) Optimizing system parameters.
For instance, by increasing the qubit-resonator coupling, the useful regime can increase
to $\sim 200\,\rm MHz$ with the $ZZ$ coupling suppressed below $20\,\rm kHz$ (see Fig.~\ref{fig16} in
Appendix~\ref{B}). In addition, decreasing the control-target detuning could be an alternative
approach, but to ensure high-fidelity gates, the detuning should be far away from the on-resonance
condition (S1). (ii) Adjusting qubit or resonator parameters after fabrication. For
example, similar to the mitigation of the frequency-collision issue, this issue can be
largely addressed by improving the precision in setting qubit frequencies with the
laser-annealing techniques \cite{Hertzberg2021,Zhang2020}. Besides, given that the frequency
reproducibility of resonators can be far better than that of qubits (see, e.g., \cite{Norris2023}),
the qubit and the resonator coupler could be placed in separate chips within flip-chip
architectures. As a consequence, each qubit can have its dedicated resonator coupler for realizing
a fast-speed gate with $ZZ$ suppression.

\section{conclusion}\label{SecV}

We introduce a CR architecture based on fixed-frequency transmons and fixed qubit-qubit couplings
for mitigating both quantum crosstalk and frequency-collision issues. Within the CR
architecture, we show that the proposed lightweight resonator coupler allows us to address the speed-fidelity
tradeoff issue imposed by quantum crosstalk and extend the usable operating region with
sizable $ZX$ couplings. Given typical qubit parameters, we demonstrate, both analytically
and numerically, that $ZZ$ quantum crosstalk can be suppressed and fast-speed, high-fidelity CR gates can
also be achieved with the condition of negative control-target detunings. 

Accordingly, we movebeyond the existing literature by operating the qubit system at both the positive and negative
detuning regions. This could largely mitigate the frequency-collision issue in existing
architectures. To assess the feasibility and utility, assuming the state-of-the-art precision
in setting frequencies, we systematically analyze the CX gate performance in the proposed
architecture and show that quantum crosstalk and frequency-collision issue can be largely mitigated, 
while the remaining collision issue can be addressed by adding weak off-resonance
drives. This suggests that the architecture proposed here could be feasible, even
considering practical challenges, especially fabrication uncertainties.

Although here we mainly focus on the heavy-hexagonal layout, one can reasonably expect that
within the introduced CR architecture, the quantum crosstalk and the frequency-collision issues
in square qubit lattices can also be suppressed. However, as studied in previous
works \cite{Hertzberg2021,Chamberland2020}, to avoid frequency collisions and to improve
the fabrication yield, more stringent requirements, especially in setting the qubit frequency, would
be required for the square layout than for the heavy-hexagonal layout.

To further improve gate performance and mitigate frequency collisions, we give an intuitive
model based on the dressed state picture. Accordingly, we illustrate the general underlying mechanism of 
frequency collisions and show that frequency collisions can be classified into two main
types, i.e., static and dynamic. We further provide dedicated
schemes to mitigate the two types of frequency collisions. While the analysis focuses on the
CR architecture with transmon qubits, we expect that these discussions could also be useful
for other qubit architecture with all-microwave-activated gate
operations \cite{Mitchell2021,Wei2022,Nesterov2021,Xiong2022}.

Supporting long coherence times and low control overhead while protecting the qubit system
from quantum crosstalk effects should be one of the most crucial steps toward large-scale
quantum processors based on fixed-frequency qubits and fixed couplings. The present
work could be helpful in guiding the design of CR gate-based architectures for this
purpose.

\begin{acknowledgments}
The author would like to thank Pei Liu, Yingshan Zhang, and Ziting Wang for many helpful discussions on the CR
gate-based transmon architecture. Thanks also go to Meng-Jun Hu, Zhikun Han, and Fei Yan for their insightful comments, especially
about the time stability of qubit performance in superconducting quantum processors. The author would also like to
thank Guangming Xue, Peng Xu, and Haifeng Yu for their generous support and encouragement. The author
gratefully acknowledges support from the National Natural Science Foundation of
China (Grant No.12204050) and the Beijing Academy
of Quantum Information Sciences.
\end{acknowledgments}

\appendix

\section{Parasitic interactions and their impact on the CR architecture}\label{A}

\begin{figure}[tbp]
\begin{center}
\includegraphics[keepaspectratio=true,width=\columnwidth]{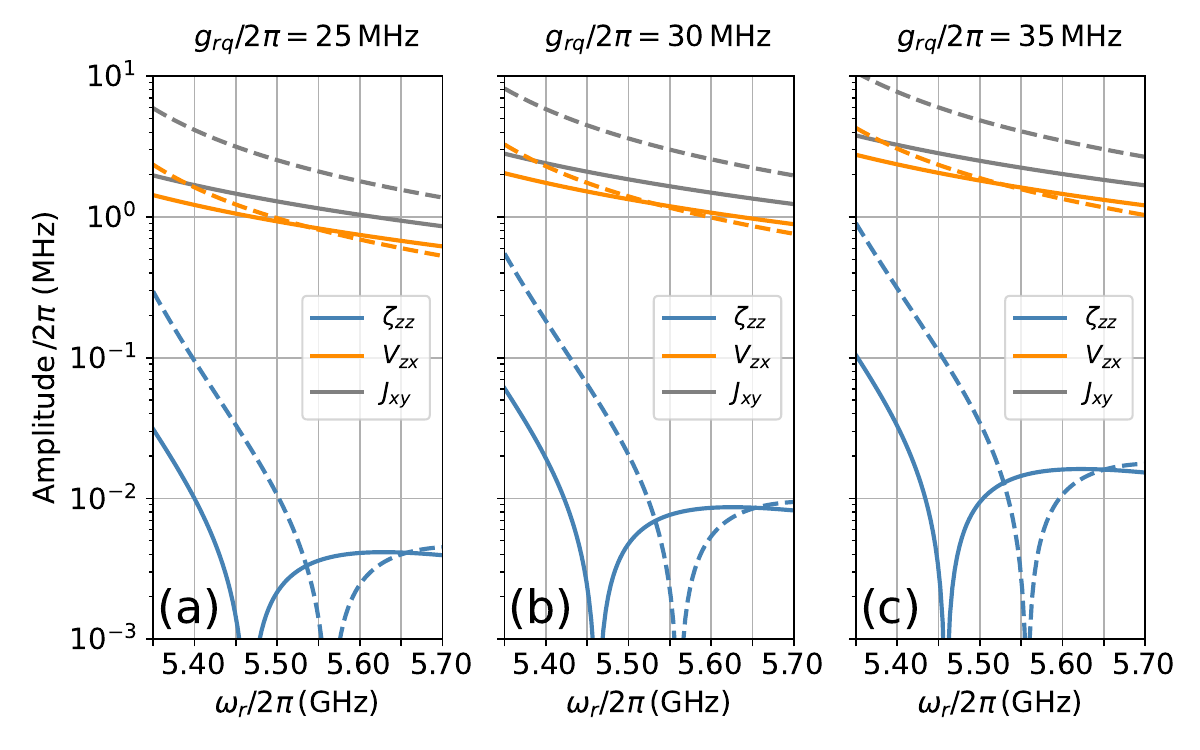}
\end{center}
\caption{Effective interqubit couplings ($ZZ$ and $XY$ couplings) and $ZX$ rates versus the resonator
frequency under the consideration of stay qubit-qubit couplings. Here, we consider the
direct qubit-qubit coupling with the strength of $g_{12}=2g_{r1}g_{r2}/\omega_{r}$.
In (a), (b), and (c), the strengths of the qubit-resonator couplings are $25\,\rm MHz$, $30\,\rm MHz$,
and $35\,\rm MHz$, respectively. The other used parameters are the same as in Fig.~\ref{fig3}.
The solid and dashed lines denote the results for the system with positive and
negative control-target detunings, respectively.}
\label{fig15}
\end{figure}

As mentioned in Sec.~\ref{SecIIA}, stray couplings between qubits are ubiquitous in real
superconducting qubit devices. Here, we turn to provide specific cases for illustrating
their effects on $ZZ$ suppression and $ZX$ rates. As mentioned in the second column of
Table~\ref{tab1:CR architectures}, we consider a direct qubit-qubit coupling arising from an
effective capacitance \cite{Ku2020,Galiautdinov2012}, giving rise
to $g_{12}=2g_{r1}g_{r2}/\omega_{r}$. By including such direct
qubit-qubit coupling terms, we go back to the generalized system model described by
the Hamiltonian in Eq.~(\ref{eq1}). Following Eqs.~(\ref{eq6}),~(\ref{eq12}), and~(\ref{eq17}), in Fig.~\ref{fig15}, we show the
$ZZ$ coupling, $XY$ coupling, and the $ZX$ rate as functions of the resonator
frequency with different the qubit-resonator couplings. Here, the used parameters are
the same as in Fig.~\ref{fig3}.

It is shown that considering the direct couplings, suppressing $ZZ$ couplings and maintaining $ZX$ couplings
with sizable strengths can still be achieved. While for the given qubit-resonator coupling, both
the $XY$ ($ZZ$) coupling and the $ZX$ rate strength are reduced compared to that without taking into account
the stay coupling (see Figs.~\ref{fig15}(a) and~\ref{fig3}), increasing the qubit-resonator
coupling here should be a practical solution to such concern, as shown in Fig.~\ref{fig15}(b).
Moreover, the usable frequency range is comparable to that shown in Fig.~\ref{fig3}.

\section{Effect of the increased qubit-bus coupling on the CR architecture}\label{B}

\begin{figure}[tbp]
\begin{center}
\includegraphics[keepaspectratio=true,width=\columnwidth]{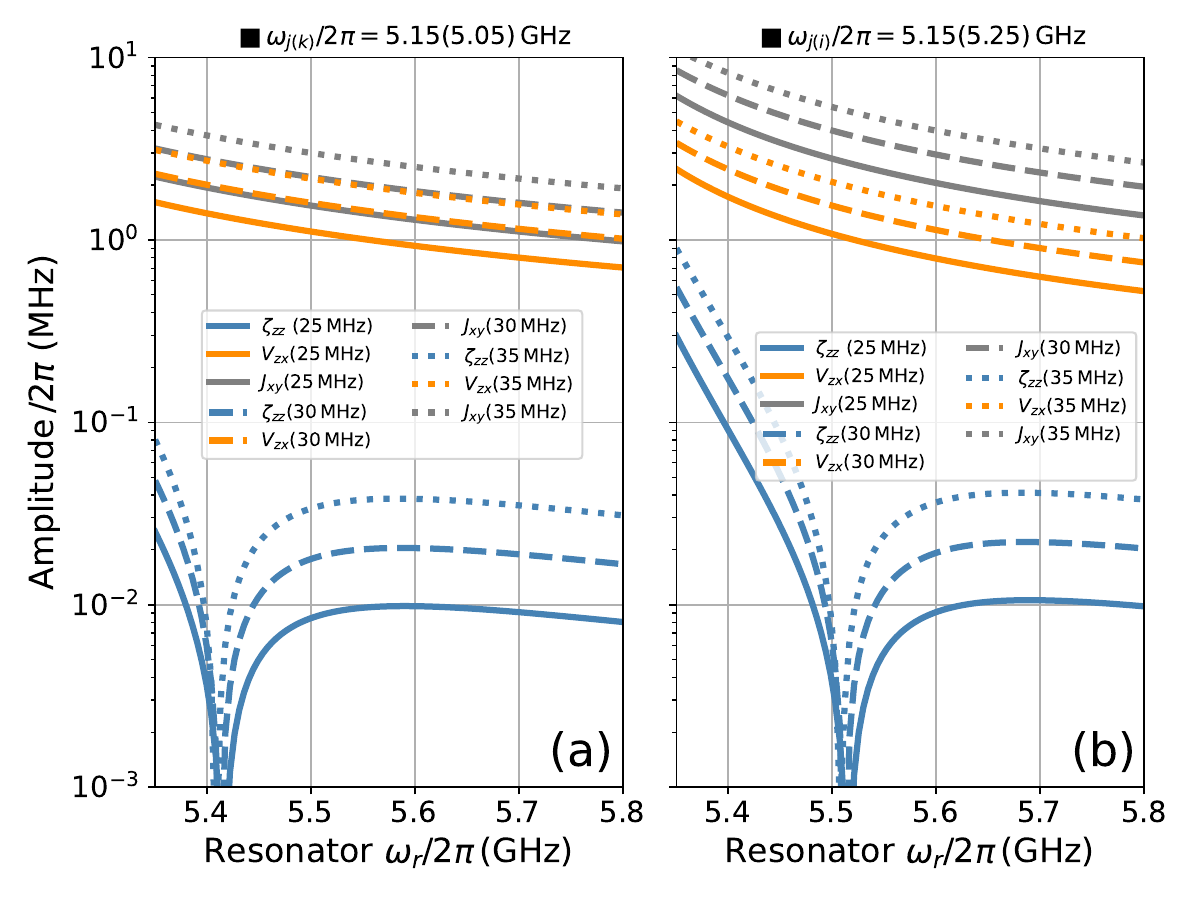}
\end{center}
\caption{Effective interqubit couplings ($ZZ$ and $XY$ couplings) and $ZX$ rates versus the resonator
frequency with different qubit-resonator couplings. (a) Positive control-target detunings.
(b) Negative control-target detunings. The solid, dashed, and dotted lines denote the results with the
qubit-resonator couplings of $25\,\rm MHz$, $30\,\rm MHz$, and $35\,\rm MHz$, respectively. The other used
parameters are the same as in Fig.~\ref{fig3}.}
\label{fig16}
\end{figure}

Within the proposed CR architecture, as mentioned in Sec.~\ref{SecIIB}, increasing
the qubit-resonator coupling could be a practical solution to further improve
the $ZX$ rate, as well as the usable parameter range. Meanwhile, this can also give rise to
larger $ZZ$ couplings. Figure~\ref{fig16} shows the effective interqubit couplings ($ZZ$ and $XY$ couplings)
and $ZX$ rates versus the resonator frequency with different qubit-resonator couplings.
It is shown that when increasing the qubit-resonator coupling, both the $ZX$ rate
and the useful range of the qubit-resonator detuning can also be enhanced.
However, for instance, increasing the qubit-resonator coupling
from $25\,\rm MHz$ to $30\,(35)\,\rm MHz$, accordingly, will increase
the typical $ZZ$ coupling from $10\,\rm kHz$ to $20\,(40)\,\rm kHz$. On the
other hand, similar to the multipath coupler \cite{Kandala2021,Zhao2021},
even in these cases, the typical magnitude of $J_{xy}/\zeta_{zz}$ can still reach the
order of $10^{2}$, suggesting that the proposed architecture can still outperform other
existing architecture, such as architectures with capacitor couplers or resonator
couplers [see Figs.~\ref{fig2}(a) and~\ref{fig2}(b)].

\section{The tune-up procedure and characterization of the direct CX gate}\label{C}

In the present work, we consider the implementation of the direct CX gate by simultaneously driving both the
control qubit $Q_{1}$ and the target qubit $Q_{2}$, as described by the drive Hamiltonian given in Eq.~({\ref{eq19}}). As mentioned
in Sec.~\ref{SecIIB}, the used pulse shape is a cosine-decorated square pulse with fixed ramp times
of $20\,\rm ns$, given in Eq.~({\ref{eq20}}). For tuning up CX gates with a fixed gate length of $t_g$, we
consider numerically optimizing the two drive amplitudes $\{\Omega_{1},\,\Omega_{2}\}$, which are the peak
drive amplitudes of the two applied drives [see Eq.~({\ref{eq20}})], and the used loss function is defined as
\begin{equation}
\begin{aligned}\label{eqC1}
L(\Omega_{1},\Omega_{2})=\langle N_{00,01}\rangle+\langle1-N_{10,11}\rangle,
\end{aligned}
\end{equation}
where $N_{00,01}$ ($N_{10,11}$) denotes the population in state $|\overline{001}\rangle$ ($|\overline{101}\rangle$) after the
gate operation with the system initialized in state $|\overline{000}\rangle$ ($|\overline{100}\rangle$).
Here, $|\overline{nml}\rangle$ denotes the eigenstate of the static Hamiltonian in Eq.~(\ref{eq14}), which is adiabatically
connected to the bare state $|nml\rangle$. The first term
in Eq.~(\ref{eqC1}) has the effect of minimizing the excitation of $Q_{2}$ when $Q_{1}$ is in
state $|0\rangle$, while the second term is for ensuring a complete flip of $Q_{2}$ when $Q_{1}$ is
in state $|1\rangle$.

After obtaining the optimal drive amplitudes $\{\Omega_{1},\,\Omega_{2}\}$, according to the full
system Hamiltonian, see Eqs.~({\ref{eq14}}) and~({\ref{eq19}}),
\begin{equation}
\begin{aligned}\label{eqC2}
H_{full}=H_{static}+H_{drive},
\end{aligned}
\end{equation}
and the pulse shape given in Eq.~({\ref{eq20}}), the actual evolution operator is given by
\begin{equation}
\begin{aligned}\label{eqC3}
U=\mathcal{\hat{T}} {\rm exp}\left(-i\int_{0}^{t_g}H_{full}(t)dt\right),
\end{aligned}
\end{equation}
where $\mathcal{\hat{T}}$ denotes the time-ordering operator. Up to single-qubit Z phases, the
gate fidelity of the implemented CX gate is \cite{Pedersen2007}
\begin{equation}
\begin{aligned}\label{eqC4}
F=\frac{{\rm Tr}(\tilde{U}^{\dagger}\tilde{U})+|{\rm Tr}(U_{\rm CX}^{\dag}\tilde{U})|^{2}}{20},
\end{aligned}
\end{equation}
where $\tilde{U}$ denotes the actual evolution operator, which is truncated to the two-qubit
computational subspace spanned by $\{|\overline{000}\rangle,|\overline{001}\rangle,|\overline{100}\rangle,|\overline{101}\rangle\}$
and $U_{\rm cx}$ denote the target CX gate
\begin{equation}
U_{\rm CX}=\left(
\begin{array}{cccc}
1 & 0 & 0& 0 \\
0 & 1 & 0 & 0 \\
0 & 0 & 0 & 1\\
0 & 0 & 1 & 0\\
\end{array}
\right).
\end{equation}

In our analysis of the CX gate performance, the above procedure is applied to both
the isolated two-qubit system shown in Fig.~{\ref{fig4}} and the four-qubit systems
shown in Fig.~{\ref{fig5}}. Additionally, for tuning up and characterizing an isolated CX gate
in the multiqubit system, we always assume that all the spectator qubits are in their ground
states. Then, the simultaneous gates are characterized based on the pulse parameters
obtained from the tune-up procedure of the constituent isolated gates. Lastly, note here that
in our numerical analysis, each transmon qubit is modeled as a four-level anharmonic oscillator
and the resonator is truncated to the lowest four energy levels.

\begin{figure}[tbp]
\begin{center}
\includegraphics[keepaspectratio=true,width=\columnwidth]{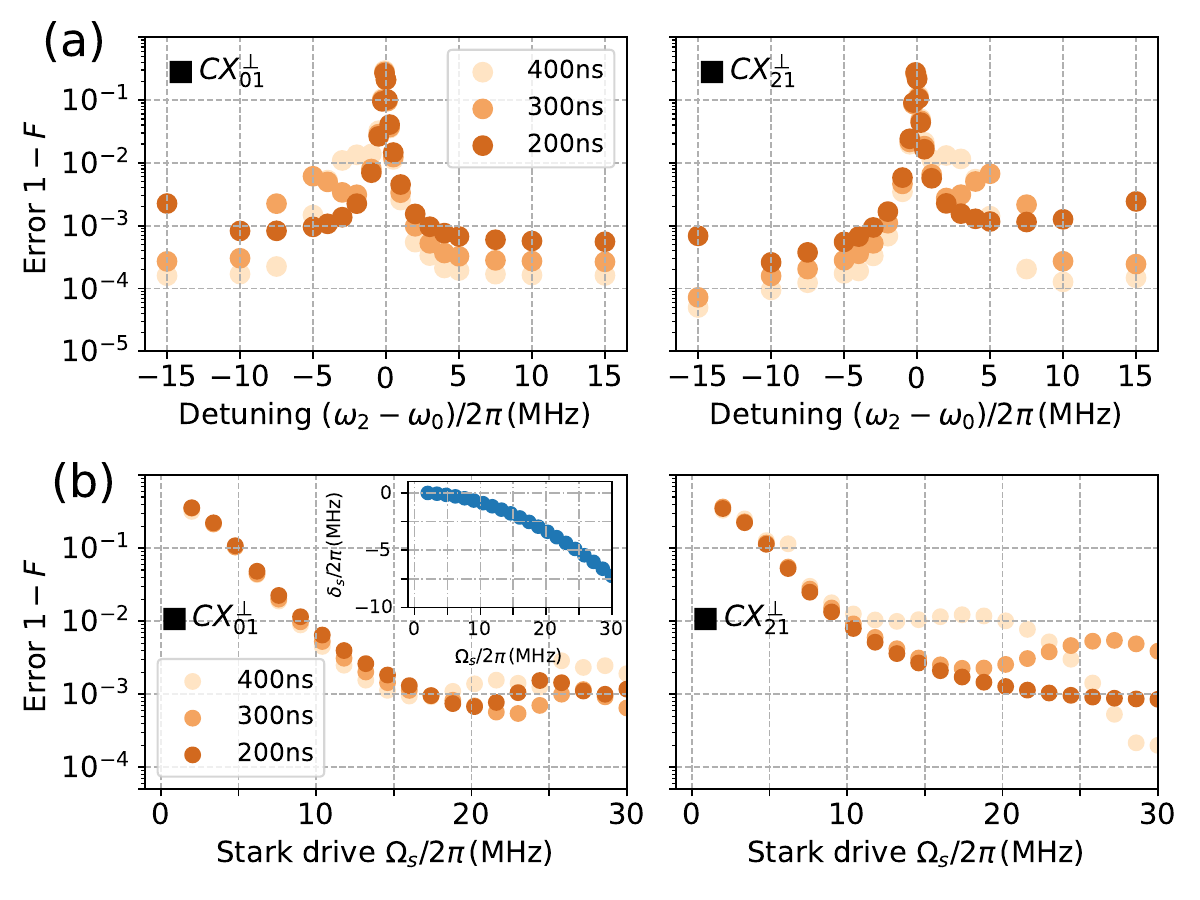}
\end{center}
\caption{(a) CX gate errors versus the detuning between neighboring control qubits ($Q_{0}$ and $Q_{2}$) with different gate
lengths for the subsystem of the '$\bot$'-shape system. Here, the qubit frequencies of $Q_{0}$ and $Q_{1}$
are $5.15\,\rm GHz$ and $5.25\,\rm GHz$, respectively. The other parameters are the same as
in Figs.~\ref{fig11}(g) and~\ref{fig11}(h). (b) CX gate errors with the presence of an always-on off-resonance
drive applied to $Q_{0}$. Here, the frequency of $Q_{2}$ is $5.25\,\rm GHz$ and the Stark drive
frequency is $5.20\,\rm GHz$. The inset shows the ac-Stark shift $\delta_{s}$ versus the drive amplitude $\Omega_{s}$.}
\label{fig17}
\end{figure}

\section{Mitigating the frequency collision issue with ac-Stark shift}\label{D}

\begin{figure}[tbp]
\begin{center}
\includegraphics[keepaspectratio=true,width=\columnwidth]{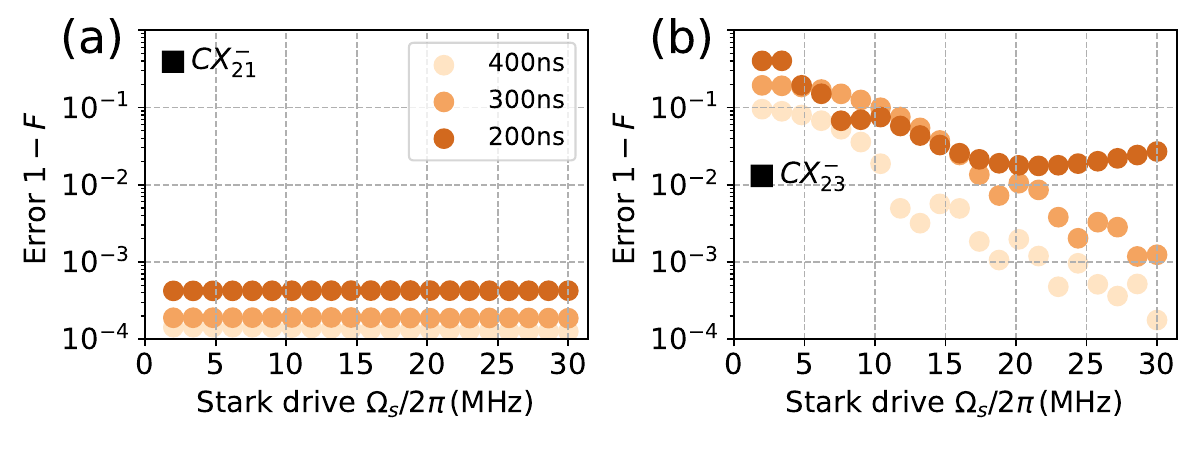}
\end{center}
\caption{CX gate errors versus the stark drive amplitude with different gate
lengths for the subsystem of the '$-$'-shape system. Here, the qubit frequencies of $Q_{1}$ and $Q_{2}$
are $5.05\,\rm GHz$ and $5.15\,\rm GHz$, respectively, while the $Q_{3}$'s frequency 
are $5.28\,\rm GHz$, $5.279\,\rm GHz$, and $5.2745\,\rm GHz$ for the gate lengths
of $400\,\rm ns$, $300\,\rm ns$, and $200\,\rm ns$, respectively (as here the collision condition
depends on the drive amplitude). The Stark drive is applied to $Q_{3}$ at the frequency of $5.33\,\rm GHz$
and the other parameters are the same as in Figs.~\ref{fig11}(c).}
\label{fig18}
\end{figure}

As illustrated in Sec.~\ref{SecIIIB}, the static collision arising from weak qubit-qubit
couplings can be avoided by adding weak off-resonance Stark drives. Here, we provide
additional illustrations on this subject. Similar to 
Fig.~{\ref{fig12}}(a), Figure~{\ref{fig17}}(a) shows that the static collision
arising from the always-on weak coupling between neighboring control qubits $Q_{0}$
and $Q_{2}$ can also exist for systems with negative control-target detunings. By applying an
off-resonance drive to $Q_{0}$ at the frequency of $5.20\,\rm GHz$, this collision issue can
also be mitigated by employing the ac-Stark effect, see Fig.~{\ref{fig17}}(b).

In addition, while in Sec.~\ref{SecIIIB}, the dynamic collision issue associated
with the resonator is addressed at the device level, i.e., by increasing the resonator
frequency, here we further show that this issue can also be mitigated by adding
off-resonance drive to selectly shift qubit frequencies. Accordingly, Figure~{\ref{fig18}}
shows the CX gate performance in the subsystem (comprising $Q_{1}$, $Q_{2}$, and $Q_{3}$) of
the '$-$'-shape four-qubit system with the off-resonance drive applied to $Q_{3}$. It is
shown that the gate performance can indeed be improved. By optimizing the gate length and
the Stark drive parameters, we expect that even higher gate fidelities can be achieved.

Note here that besides two-qubit gates, the impact of the introduced
off-resonance drive on single-qubit gates should also be examined. As demonstrated theoretically in
the previous work \cite{Zhao2022c}, even in the presence of off-resonance drives, single-qubit gates with gate errors
below $0.0001$ could still be achieved, especially when using a small drive amplitude, as in the present work.

\section{Newly added frequency collisions}\label{E}

Within the currently studied architecture, the magnitude of the qubit-resonator detuning is
comparable to that of the control-target detuning and the qubit anharmonicity. This
can thus introduce new frequency-collision issues associated with both the qubits and
the resonators. Moreover, since the resonator frequency is larger than the qubit
frequency, with the typical detuning of $250\,\rm MHz$, frequency-collision issues similar
to (S1-S2) and (D1-D4) (see Sec.~\ref{SecIIA}) can be ignored here. Considering
that the resonator couplers are generally initialized in their ground
states, the leading frequency collision should be manifested as CR drive-induced
two-photon transitions.

To be more specific, we consider a coupled qubit-resonator system, wherein $\Delta_{q}=\omega_{q}-\omega_{r}$
and $g_{rq}$ denote the qubit-resonator detuning and the qubit-resonator coupling,
respectively, and the system state is labeled by $|Q,R\rangle$. We further assume
that a drive with frequency $\omega_{d}$ and drive magnitude 
$\Omega$ is applied to the qubit, which is a transmon qubit with 
anharmonicity $\alpha_{q}$. Within such a coupled qubit-resonator
system, there are two main types of two-photon transitions with the resonance
conditions
\begin{equation}
\begin{aligned}\label{eqE1}
&\omega_{r}+\omega_{q}=2\omega_{d},
\\&\omega_{r}+\omega_{q}+\alpha_{q}=2\omega_{d},
\end{aligned}
\end{equation}
which correspond to the two-photon transition $|00\rangle\leftrightarrow|11\rangle$
and $|10\rangle\leftrightarrow|21\rangle$, respectively. By applying second-order perturbation
theory \cite{Nesterov2021,Tripathi2019}, the transition rate can be approximated by
\begin{equation}
\begin{aligned}\label{eqE2}
&V_{00\leftrightarrow11}\simeq\frac{2g_{rq}\Omega^{2}\alpha_{q}}{(\alpha_{q}+\Delta_{q})\Delta_{q}^{2}},
\\&V_{10\leftrightarrow21}\simeq\frac{-\sqrt{2}\Omega}{\Delta_{q}+\alpha_{q}}
\big[\frac{-4g_{rq}\Omega}{\Delta_{q}+\alpha_{q}}+\frac{g_{rq}\Omega}{\Delta_{q}}+\frac{3g_{rq}\Omega}{\Delta_{q}+2\alpha_{q}} \big].
\end{aligned}
\end{equation}

As mentioned in Sec.~\ref{SecIII}, considering the three-frequency
setting, i.e., $\omega_{i}>\omega_{j}>\omega_{k}$, for the heavy-hexagonal qubit lattice and the typical
parameters listed in Table~\ref{tab2:System parameters}, we find that, within the currently studied
CR architecture, the dominant frequency collision should be the two-photon
transition $|00\rangle\leftrightarrow|11\rangle$ and the collision condition is given by
\begin{equation}
\begin{aligned}\label{eq23}
\omega_{r}+\omega_{j}=2\omega_{i}.
\end{aligned}
\end{equation}
On the other hand, as indicated by Eq.~(\ref{eqE2}), the two-photon
transition $|10\rangle\leftrightarrow|21\rangle$ is heavily suppressed by the large
qubit-resonator detuning and the negative anharmonicity of transmon qubits.


\begin{thebibliography}{99}

\bibitem{Martinis2015} J. M. Martinis, Qubit Metrology for Building a Fault Tolerant Quantum Computer, \href{https://doi.org/10.1038/npjqi.2015.5}{npj Quantum Inf. \textbf{1}, 15005 (2015)}.

\bibitem{Krantz2019} P. Krantz, M. Kjaergaard, F. Yan, T. P. Orlando, S. Gustavsson, and W. D. Oliver, A Quantum Engineer's Guide to Superconducting Qubits, \href{https://doi.org/10.1063/1.5089550}{Appl. Phys. Rev. \textbf{6}, 021318 (2019)}.

\bibitem{Blais2021} A. Blais, A. L. Grimsmo, S. M. Girvin, and A. Wallraff, Circuit quantum electrodynamics, \href{https://doi.org/10.1103/RevModPhys.93.025005}{Rev. Mod. Phys. \textbf{93}, 025005 (2021)}.

\bibitem{Bialczak2011} R. C. Bialczak, M. Ansmann, M. Hofheinz, M. Lenander, E. Lucero, M. Neeley, A. D. O'Connell, D. Sank, H. Wang, M. Weides, J. Wenner, T. Yamamoto, A. N. Cleland, and J. M. Martinis, Fast Tunable Coupler for Superconducting Qubits, \href{https://doi.org/10.1103/PhysRevLett.106.060501}{Phys. Rev. Lett. \textbf{106}, 060501 (2011)}.

\bibitem{Paraoanu2006} G. S. Paraoanu, Microwave-induced coupling of superconducting qubits
, \href{https://doi.org/10.1103/PhysRevB.74.140504}{Phys. Rev. B \textbf{74}, 140504(R) (2006)}.

\bibitem{Rigetti2010} C. Rigetti and M. Devoret, Fully microwave-tunable universal gates in superconducting qubits with linear couplings and fixed transition frequencies, \href{https://doi.org/10.1103/PhysRevB.81.134507}{Phys. Rev. B \textbf{81}, 134507 (2010)}.

\bibitem{de Groot2010} P. C. de Groot, J. Lisenfeld, R. N. Schouten, S. Ashhab, A. Lupa\c{s}cu, C. J. P. M. Harmans, and J. E. Mooij, Selective darkening of degenerate transitions demonstrated with two superconducting quantum bits, \href{https://doi.org/10.1038/nphys1733}{Nat. Phys. \textbf{6}, 763 (2010)}.

\bibitem{Chow2011} J. M. Chow, A. D. C\'{o}rcoles, J. M. Gambetta, C. Rigetti, B. R. Johnson, J. A. Smolin, J. R. Rozen, G. A. Keefe, M. B. Rothwell, M. B. Ketchen, and M. Steffen, Simple All-Microwave Entangling Gate for Fixed-Frequency Superconducting Qubits, \href{https://doi.org/10.1103/PhysRevLett.107.080502}{Phys. Rev. Lett. \textbf{107}, 080502 (2011)}.

\bibitem{Koch2007} J. Koch, T. M. Yu, J. Gambetta, A. A. Houck, D. I. Schuster, J. Majer, A. Blais, M. H. Devoret, S.
    M. Girvin, and R. J. Schoelkopf, Charge-insensitive qubit design derived from the cooper pair box, \href{https://doi.org/10.1103/PhysRevA.76.042319}{Phys. Rev. A \textbf{76}, 042319 (2007)}.

\bibitem{Chen2022} E. H. Chen, T. J. Yoder, Y. Kim, N. Sundaresan, S. Srinivasan, M. Li, A. D. C\'{o}rcoles, A. W. Cross, and M. Takita, Calibrated Decoders for Experimental Quantum Error Correction, \href{https://doi.org/10.1103/PhysRevLett.128.110504}{Phys. Rev. Lett. \textbf{128}, 110504 (2022)}.

\bibitem{Sundaresan2023} N. Sundaresan, T. J. Yoder, Y. Kim, M. Li, E. H. Chen, G. Harper, T. Thorbeck, A. W. Cross, A. D. C\'{o}rcoles, and M. Takita, Demonstrating multi-round subsystem quantum error correction using matching and maximum likelihood decoders, \href{https://doi.org/10.1038/s41467-023-38247-5}{Nat. Commun. \textbf{14}, 2852 (2023)}.

\bibitem{Kim2023} Y. Kim, A. Eddins, S. Anand, K. X. Wei, E. v. d. Berg, S. Rosenblatt, H. Nayfeh, Y. Wu, M. Zaletel, K. Temme, and A. Kandala, Evidence for the utility of quantum computing before fault tolerance, \href{https://doi.org/10.1038/s41586-023-06096-3}{Nature \textbf{618}, 500-505 (2023)}.

\bibitem{Jurcevic2021} P. Jurcevic, A. Javadi-Abhari, L. S. Bishop, I. Lauer, D. F. Bogorin, M. Brink, L. Capelluto, O. G\"{u}nl\"{u}k, T. Itoko, N. Kanazawa \emph{et al}., Demonstration of quantum volume 64 on a superconducting quantum computing system, \href{https://doi.org/10.1088/2058-9565/abe519}{Quantum Sci. Technol. \textbf{6}, 025020 (2021)}.

\bibitem{Chow2012} J. M. Chow, J. M. Gambetta, A. D. C\'{o}rcoles, S. T. Merkel, J. A. Smolin, C. Rigetti, S. Poletto, G. A. Keefe, M. B. Rothwell, J. R. Rozen, M. B. Ketchen, and M. Steffen, Universal Quantum Gate Set Approaching Fault-Tolerant Thresholds with Superconducting Qubits, \href{https://doi.org/10.1103/PhysRevLett.109.060501}{Phys. Rev. Lett. \textbf{109}, 060501 (2012)}.

\bibitem{Place2021} A. P. M. Place, L. V. H. Rodgers, P. Mundada, B. M. Smitham, M. Fitzpatrick, Z. Leng, A. Premkumar, J. Bryon, A. Vrajitoarea, S. Sussman \emph{et al.}, New material platform for superconducting transmon qubits with coherence times exceeding 0.3 milliseconds, \href{https://doi.org/10.1038/s41467-021-22030-5}{Nat. Commun. \textbf{12}, 1779 (2021)}.

\bibitem{Wang2022} C. Wang, X. Li, H. Xu, Z. Li, J. Wang, Z. Yang, Z. Mi, X. Liang, T. Su, C. Yang \emph{et al.}, Towards practical quantum computers: transmon qubit with a lifetime approaching 0.5 milliseconds, \href{https://doi.org/10.1038/s41534-021-00510-2}{npj Quantum Inf. \textbf{8}, 3 (2022)}.

\bibitem{Gordon2022} R. T. Gordon, C. E. Murray, C. Kurter, M. Sandberg, S. A. Hall, K. Balakrishnan, R. Shelby, B. Wacaser, A. A. Stabile, J.W. Sleight \emph{et al.}, Environmental Radiation Impact on Lifetimes and Quasiparticle Tunneling Rates of Fixed-Frequency Transmon Qubits, \href{https://aip.scitation.org/doi/10.1063/5.0078785}{Appl. Phys. Lett. \textbf{120}, 074002 (2022)}.

\bibitem{Sheldon2016} S. Sheldon, E. Magesan, J. M. Chow, and J. M. Gambetta, Procedure for systematically tuning up cross-talk in the cross-resonance gate, \href{https://doi.org/10.1103/PhysRevA.93.060302}{Phys. Rev. A \textbf{93}, 060302(R) (2016)}.

\bibitem{Patterson2019} A. D. Patterson, J. Rahamim, T. Tsunoda, P. A. Spring, S. Jebari, K. Ratter, M. Mergenthaler, G. Tancredi, B. Vlastakis, M. Esposito, and P. J. Leek, Calibration of a Cross-Resonance Two-Qubit Gate Between Directly Coupled Transmons, \href{https://doi.org/10.1103/PhysRevApplied.12.064013}{Phys. Rev. Appl. \textbf{12}, 064013 (2019)}.

\bibitem{Paik2020} H. Paik, S. Srinivasan, S. Rosenblatt, J. Chavez-Garcia, D. Bogorin, O. Jinka, G. Keefe, D. Shao, J.-B. Yau, M. Brink, and J. M. Chow, Coupler characterization of superconducting transmons qubits for cross-resonance gate, \href{https://doi.org/10.1109/IEDM13553.2020.9371955}{2020 IEEE International Electron Devices Meeting (IEDM), 2020, pp. 38.2.1-38.2.4}.

\bibitem{Chow2015} J. M. Chow, S. J. Srinivasan, E. Magesan, A. D. C\'{o}rcoles, D. W. Abraham, J. M. Gambetta, and M. Steffen, Characterizing a four-qubit planar lattice for arbitrary error detection, \href{https://doi.org/10.1117/12.2192740}{Proc. SPIE 9500, Quantum Inf. Comput. \textbf{13}, 95001G (2015)}.

\bibitem{Gambetta2017} J. M. Gambetta, J. M. Chow, and M. Steffen, Building logical qubits in a superconducting quantum computing system, \href{https://doi.org/10.1038/s41534-016-0004-0}{npj Quantum Inf. \textbf{3}, 2 (2017)}.

\bibitem{Brink2018} M. Brink, J. M. Chow, J. Hertzberg, E. Magesan, and Sami Rosenblatt, Device challenges for near term superconducting quantum processors: frequency collisions, \href{https://doi.org/10.1109/IEDM.2018.8614500}{2018 IEEE International Electron Devices Meeting (IEDM), (2018)}.

\bibitem{Hertzberg2021} J. B. Hertzberg, E. J. Zhang, S. Rosenblatt, E. Magesan, J. A. Smolin, J.-B. Yau, V. P. Adiga, M. Sandberg, M. Brink, Je. M. Chow, and J. S. Orcutt, Laser-annealing Josephson junctions for yielding scaled-up superconducting quantum processors, \href{https://doi.org/10.1038/s41534-021-00464-5}{npj Quantum Inf. \textbf{7}, 129 (2021)}.

\bibitem{Kreikebaum2020} J. M. Kreikebaum, K. P. O'Brien, A. Morvan, and I. Siddiqi, Improving wafer-scale Josephson junction resistance
variation in superconducting quantum coherent circuits, \href{https://doi.org/10.1088/1361-6668/ab8617}{Supercond. Sci. Technol. \textbf{33}, 06LT02 (2020)}.

\bibitem{DiCarlo2009} L. DiCarlo, J. M. Chow, J. M. Gambetta, L. S. Bishop, B. R. Johnson, D. I. Schuster, J. Majer, A. Blais, L. Frunzio, S. M. Girvin, and R. J. Schoelkopf, Demonstration of two-qubit algorithms with a superconducting quantum processor, \href{https://doi.org/10.1038/nature08121}{Nature (London) \textbf{460}, 240 (2009)}.

\bibitem{Gambetta2012a} J. M. Gambetta, A. D. C\'{o}rcoles, S. T. Merkel, B. R. Johnson, John A. Smolin, J. M. Chow, C. A. Ryan, C. Rigetti, S. Poletto, T. A. Ohki, M. B. Ketchen, and M. Steffen, Characterization of Addressability by Simultaneous Randomized Benchmarking
, \href{https://doi.org/10.1103/PhysRevLett.109.240504}{Phys. Rev. Lett. \textbf{109}, 240504 (2012)}.

\bibitem{McKay2019} D. C. McKay, S. Sheldon, J. A. Smolin, J. M. Chow, and J. M. Gambetta, Three Qubit Randomized Benchmarking, \href{https://doi.org/10.1103/PhysRevLett.122.200502}{Phys. Rev. Lett. \textbf{122}, 200502 (2019)}.


\bibitem{Malekakhlagh2020} M. Malekakhlagh, E. Magesan, and D. C. McKay, First-principles analysis of cross-resonance gate operation, \href{https://doi.org/10.1103/PhysRevA.102.042605}{Phys. Rev. A \textbf{102}, 042605 (2020)}.

\bibitem{Cai2021}  T.-Q. Cai, X.-Y. Han, Y.-K. Wu, Y.-L. Ma, J.-H. Wang, Z.-L. Wang, H.-Y Zhang, H.-Y Wang, Y.-P. Song, and L.-M. Duan, Impact of Spectators on a Two-Qubit Gate in a Tunable Coupling Superconducting Circuit, \href{https://doi.org/10.1103/PhysRevLett.127.060505}{Phys. Rev. Lett. \textbf{127}, 060505 (2021)}.

\bibitem{Heya2023} K. Heya, M. Malekakhlagh, S. Merkel, N. Kanazawa, and E. Pritchett, Floquet Analysis of Frequency Collisions, \href{https://doi.org/10.48550/arXiv.2302.12816}{arXiv:2302.12816}.

\bibitem{Morvan2022} A. Morvan, L. Chen, J. M. Larson, D. I. Santiago, and I. Siddiqi, Optimizing frequency allocation for fixed-frequency superconducting quantum processors, \href{https://doi.org/10.1103/PhysRevResearch.4.023079}{Phys. Rev. Research \textbf{4}, 023079 (2022)}.

\bibitem{Chamberland2020} C. Chamberland, G. Zhu, T. J. Yoder, J. B. Hertzberg, and A. W. Cross, Topological and Subsystem Codes on Low-Degree Graphs with Flag Qubits, \href{https://doi.org/10.1103/PhysRevX.10.011022}{Phys. Rev. X \textbf{10}, 011022 (2020)}.

\bibitem{Ku2020} J. Ku, X. Xu, M. Brink, D. C. McKay, J. B. Hertzberg, M. H. Ansari, and B. L. T. Plourde, Suppression of Unwanted ZZ Interactions in a Hybrid Two-Qubit System, \href{https://doi.org/10.1103/PhysRevLett.125.200504}{Phys. Rev. Lett. \textbf{125}, 200504 (2020)}.

\bibitem{Zhao2020} P. Zhao, P. Xu, D. Lan, J. Chu, X. Tan, H. Yu, and Y. Yu, High-Contrast ZZ Interaction Using Superconducting Qubits with Opposite-Sign Anharmonicity, \href{https://doi.org/10.1103/PhysRevLett.125.200503}{Phys. Rev. Lett. \textbf{125}, 200503 (2020)}.

\bibitem{Xu2021} X. Xu and M.H. Ansari, ZZ freedom in two qubit gates, \href{https://doi.org/10.1103/PhysRevApplied.15.064074}{Phys. Rev. Appl. \textbf{15}, 064074 (2021)}.

\bibitem{Chavez-Garcia2022} J. M. Ch\'{a}vez-Garcia, F. Solgun, J. B. Hertzberg, O. Jinka, M. Brink, and B. Abdo, Weakly Flux-Tunable Superconducting Qubit, \href{https://doi.org/10.1103/PhysRevApplied.18.034057}{Phys. Rev. Appl. \textbf{18}, 034057 (2022)}.

\bibitem{Xu2023} Xuexin Xu and M. Ansari, Parasitic-Free Gate: An Error-Protected Cross-Resonance Switch in Weakly Tunable Architectures, \href{https://doi.org/10.1103/PhysRevApplied.19.024057}{Phys. Rev. Appl. \textbf{19}, 024057 (2023)}.

\bibitem{Sundaresan2020} N. Sundaresan, I. Lauer, E. Pritchett, E. Magesan, P. Jurcevic, and J. M. Gambetta, Reducing Unitary and Spectator Errors in Cross Resonance with Optimized Rotary Echoes, \href{https://doi.org/10.1103/PRXQuantum.1.020318}{PRX Quantum \textbf{1}, 020318 (2020)}.

\bibitem{Mitchell2021} B. K. Mitchell, R. K. Naik, A. Morvan, A. Hashim, J. M. Kreikebaum, B. Marinelli, W. Lavrijsen, K. Nowrouzi, D. I. Santiago, and I. Siddiqi, Hardware-Efficient Microwave-Activated Tunable Coupling Between Superconducting Qubits, \href{https://doi.org/10.1103/PhysRevLett.127.200502}{Phys. Rev. Lett. \textbf{127}, 200502 (2021)}.

\bibitem{Wei2022} K. X. Wei, E. Magesan, I. Lauer, S. Srinivasan, D. F. Bogorin, S. Carnevale, G. A. Keefe, Y. Kim, D. Klaus, W. Landers, N. Sundaresan, C. Wang, E. J. Zhang, M. Steffen, O. E. Dial, D. C. McKay, and A. Kandala, Hamiltonian Engineering with Multicolor Drives for Fast Entangling Gates
and Quantum Crosstalk Cancellation, \href{https://doi.org/10.1103/PhysRevLett.129.060501}{Phys. Rev. Lett. \textbf{129}, 060501 (2022)}.

\bibitem{Kandala2021} A. Kandala, K. X. Wei, S. Srinivasan, E. Magesan, S. Carnevale, G. A. Keefe, D. Klaus, O. Dial, and D. C. McKay, Demonstration of a High-Fidelity cnot Gate for Fixed-Frequency Transmons with Engineered ZZ Suppression, \href{https://doi.org/10.1103/PhysRevLett.127.130501}{Phys. Rev. Lett. \textbf{127}, 130501 (2021)}.

\bibitem{Zhao2021} P. Zhao, D. Lan, P. Xu, G. Xue, M. Blank, X. Tan, H. Yu, and Y. Yu, Suppression of Static ZZ Interaction in an All-Transmon Quantum Processor, \href{https://doi.org/10.1103/PhysRevApplied.16.024037}{Phys. Rev. Appl. \textbf{16}, 024037 (2021)}.

\bibitem{Goerz2017} M. H. Goerz, F. Motzoi, K. B. Whaley, and C. P. Koch, Charting the circuit QED design landscape using optimal control theory, \href{https://doi.org/10.1038/s41534-017-0036-0}{npj Quantum Inf. \textbf{3}, 37 (2017)}.

\bibitem{Jin2021} L. Jin, Implementing High-fidelity Two-Qubit Gates in Superconducting Coupler Architecture with Novel Parameter Regions, \href{https://arxiv.org/abs/2105.13306}{arXiv:2105.13306}.

\bibitem{Li2022} B. Li, T. Calarco, and F. Motzoi, Nonperturbative Analytical Diagonalization of Hamiltonians with Application to
Circuit QED, \href{http://dx.doi.org/10.1103/PRXQuantum.3.030313}{PRX Quantum \textbf{3}, 030313 (2022)}.

\bibitem{Zhao2022b} P. Zhao, Y. Zhang, G. Xue, Y. Jin, and H. Yu, Tunable coupling of widely separated superconducting qubits: A possible application toward a modular quantum device, \href{https://doi.org/10.1063/5.0097521}{Appl. Phys. Lett. \textbf{121}, 032601 (2022)}.

\bibitem{Tripathi2019} V. Tripathi, M. Khezri, and A. N. Korotkov, Operation and intrinsic error budget of a two-qubit cross-resonance gate, \href{https://doi.org/10.1103/PhysRevA.100.012301}{Phys. Rev. A \textbf{100}, 012301 (2019)}.

\bibitem{Ware2019} M. Ware, B. R. Johnson, J. M. Gambetta, T. A. Ohki, J. M. Chow, and B.L.T. Plourde, Cross-resonance interactions between superconducting qubits with variable detuning, \href{https://doi.org/10.48550/arXiv.1905.11480}{arXiv:1905.11480}.

\bibitem{Magesan2020} E. Magesan and J. M. Gambetta, Effective Hamiltonian models of the cross-resonance gate, \href{https://doi.org/10.1103/PhysRevA.101.052308}{Phys. Rev. A \textbf{101}, 052308 (2020)}.

\bibitem{Malekakhlagh2022} M. Malekakhlagh and E. Magesan, Mitigating off-resonant error in the cross-resonance gate, \href{https://doi.org/10.1103/PhysRevA.105.012602}{Phys. Rev. A \textbf{105}, 012602 (2022)}.


\bibitem{Noguchi2020} A. Noguchi, A. Osada, S. Masuda, S. Kono, K. Heya, S. P. Wolski, H. Takahashi, T. Sugiyama, D. Lachance-Quirion, and Y. Nakamura, Fast parametric two-qubit gates with suppressed residual interaction using the second-order nonlinearity of a cubic transmon, \href{https://journals.aps.org/pra/abstract/10.1103/PhysRevA.102.062408}{Phys. Rev. A \textbf{102}, 062408 (2020)}.

\bibitem{Xiong2022} H. Xiong, Q. Ficheux, A. Somoroff, L. B. Nguyen, E. Dogan, D. Rosenstock, C. Wang, K. N. Nesterov, M. G. Vavilov, and V. E. Manucharyan, Arbitrary controlled-phase gate on fluxonium qubits using differential ac Stark shifts, \href{https://doi.org/10.1103/PhysRevResearch.4.023040}{Phys. Rev. Research \textbf{4}, 023040 (2022)}.

\bibitem{Ni2022} Z. Ni, S. Li, L. Zhang, J. Chu, J. Niu, T. Yan, X. Deng, L. Hu, J. Li, Y. Zhong, S. Liu, F. Yan, Y. Xu, and Dapeng Yu, Scalable Method for Eliminating Residual ZZ Interaction between Superconducting Qubits, \href{https://doi.org/10.1103/PhysRevLett.129.040502}{Phys. Rev. Lett. \textbf{129}, 040502 (2022)}.

\bibitem{Zhao2022c} P. Zhao, T. Ma, Y. Jin, and H. Yu, Combating fluctuations in relaxation times of fixed-frequency transmon qubits with microwave-dressed states, \href{https://doi.org/10.1103/PhysRevA.105.062605}{Phys. Rev. A \textbf{105}, 062605 (2022)}.

\bibitem{Wang2023} Z. T. Wang, P. Zhao, Z. H. Yang, Ye Tian, H. F. Yu, and S. P. Zhao, Escaping Detrimental Interactions with Microwave-Dressed Transmon Qubits, \href{https://doi.org/10.1088/0256-307X/40/7/070304}{Chin. Phys. Lett. \textbf{40}, 070304 (2023)}.

\bibitem{Galiautdinov2012} A. Galiautdinov, A. N. Korotkov, and J. M. Martinis, Resonator-zero-qubit architecture for superconducting qubits, \href{https://doi.org/10.1103/PhysRevA.85.042321}{Phys. Rev. A \textbf{85}, 042321 (2012)}.

\bibitem{Yan2018} F. Yan, P. Krantz, Y. Sung, M. Kjaergaard, D. L. Campbell, T. P. Orlando, S. Gustavsson, and W. D. Oliver, Tunable Coupling Scheme for Implementing High-Fidelity Two-Qubit Gates, \href{https://doi.org/10.1103/PhysRevApplied.10.054062}{Phys. Rev. Appl. \textbf{10}, 054062 (2018)}.

\bibitem{Yanay2022} Y. Yanay, J. Braum\"{u}ller, T. P. Orlando, S. Gustavsson, C. Tahan, and W. D. Oliver, Mediated Interactions beyond the Nearest Neighbor in an Array of Superconducting Qubits, \href{https://doi.org/10.1103/PhysRevApplied.17.034060}{Phys. Rev. Appl. \textbf{17}, 034060 (2022)}.

\bibitem{Zhao2022a} P. Zhao, K. Linghu, Z. Li, P. Xu, R. Wang, G. Xue, Y. Jin, and H. Yu, Quantum Crosstalk Analysis for Simultaneous Gate Operations on Superconducting Qubits, \href{https://doi.org/10.1103/PRXQuantum.3.020301}{PRX Quantum \textbf{3}, 020301 (2022)}.

\bibitem{Barends2014} R. Barends, J. Kelly, A. Megrant, A. Veitia, D. Sank, E. Jeffrey, T. C. White, J. Mutus, A. G. Fowler, B. Campbell, Y. Chen, Z. Chen, B. Chiaro, A. Dunsworth, C. Neill, P. O'Malley, P. Roushan, A. Vainsencher, J. Wenner, A. N. Korotkov, A. N. Cleland, and J. M. Martinis, Superconducting quantum circuits at the surface code threshold for fault tolerance, \href{https://doi.org/10.1038/nature13171}{Nature \textbf{508}, 500 (2014)}.

\bibitem{Pedersen2007} L. H. Pedersen, N. M. M{\o}ller, and K. M{\o}lmer, Fidelity of quantum operations,
    \href{https://doi.org/10.1016/j.physleta.2007.02.069}{Phys. Lett. A \textbf{367}, 47 (2007)}.

\bibitem{Motzoi2009} F. Motzoi, J. M. Gambetta, P. Rebentrost, and F. K. Wilhelm, Simple Pulses for Elimination of Leakage in Weakly Nonlinear Qubits, \href{https://doi.org/10.1103/PhysRevLett.103.110501}{Phys. Rev. Lett. \textbf{103}, 110501 (2009)}.

\bibitem{Gambetta2011} J. M. Gambetta, F. Motzoi, S. T. Merkel, and F. K. Wilhelm, Analytic control methods for high-fidelity unitary operations in a weakly nonlinear oscillator, \href{http://dx.doi.org/10.1103/PhysRevA.83.012308}{Phys. Rev. A \textbf{83}, 012308 (2011)}.

\bibitem{Li2023} B. Li, T. Calarco, and F. Motzoi, Suppression of coherent errors in Cross-Resonance gates via recursive DRAG, \href{https://doi.org/10.48550/arXiv.2303.01427}{arXiv:2303.01427}.

\bibitem{Norris2023} G. J. Norris, L. Michaud, D. Pahl, M. Kerschbaum, C. Eichler, J.-C. Besse, and A. Wallraff, Improved Parameter Targeting in 3D-Integrated Superconducting Circuits through a Polymer Spacer Process, \href{https://doi.org/10.48550/arXiv.2307.00046}{arXiv:2307.00046}.

\bibitem{Wood2018} C. J. Wood and J. M. Gambetta, Quantification and characterization of leakage errors, \href{https://doi.org/10.1103/PhysRevA.97.032306}{Phys. Rev. A \textbf{97}, 032306 (2018)}.

\bibitem{Chu2004} S.-I. Chu and D. A. Telnov, Beyond the Floquet theorem: Generalized Floquet formalisms and quasienergy methods for atomic and molecular multiphoton processes in intense laser fields, \href{https://doi.org/10.1016/j.physrep.2003.10.001}{Phys. Rep. \textbf{390}, 1 (2004)}

\bibitem{Zhang2019} Y. Zhang, B. J. Lester, Y. Y. Gao, L. Jiang, R. J. Schoelkopf, and S. M. Girvin, Engineering bilinear mode coupling in circuit QED: Theory and experiment, \href{https://doi.org/10.1103/PhysRevA.99.012314}{Phys. Rev. A \textbf{99}, 012314 (2019)}.

\bibitem{Gandon2022} A. Gandon, C. L. Calonnec, R. Shillito, A. Petrescu, and A. Blais, Engineering, Control, and Longitudinal Readout of Floquet Qubits, \href{https://doi.org/10.1103/PhysRevApplied.17.064006}{Phys. Rev. Appl. \textbf{17}, 064006 (2022)}.


\bibitem{Schneider2018} A. Schneider, J. Braum\"{u}ller, L. Guo, P. Stehle, H. Rotzinger, M. Marthaler, A. V. Ustinov, and M. Weides, Local sensing with the multilevel ac stark effect, \href{https://journals.aps.org/pra/abstract/10.1103/PhysRevA.97.062334}{Phys. Rev. A \textbf{97}, 062334 (2018)}.


\bibitem{Zhao2023} P. Zhao, R. Wang, M.-J. Hu, T. Ma, P. Xu, Y. Jin, and H. Yu, Baseband Control of Superconducting Qubits with Shared Microwave Drives, \href{http://dx.doi.org/10.1103/PhysRevApplied.19.054050}{Phys. Rev. Appl. \textbf{19}, 054050 (2023)}.

\bibitem{Nesterov2021} K. N. Nesterov, Q. Ficheux, V. E. Manucharyan, and M. G. Vavilov, Proposal for Entangling Gates on Fluxonium Qubits via a Two-Photon Transition, \href{https://doi.org/10.1103/PRXQuantum.2.020345}{PRX Quantum \textbf{2}, 020345 (2021)}.

\bibitem{Poletto2012} S. Poletto, J. M. Gambetta, S. T. Merkel, J. A. Smolin, J. M. Chow, A. D. C\'{o}rcoles, G. A. Keefe, M. B. Rothwell, J. R. Rozen, D. W. Abraham, C. Rigetti, and M. Steffen, Entanglement of two superconducting qubits in a waveguide cavity via monochromatic two-photon excitation, \href{https://doi.org/10.1103/PhysRevLett.109.240505}{Phys. Rev. Lett. \textbf{109}, 240505 (2012)}.


\bibitem{Zhang2020} E. Zhang, S. Srinivasan, N. Sundaresan, D. F. Bogorin, Y. Martin, J. B. Hertzberg, J. Timmerwilke, E. J. Pritchett, J.-B. Yau, and C. Wang, \emph{et al.}, High-fidelity superconducting quantum processors via laser-annealing of transmon qubits, \href{https://doi.org/10.1126/sciadv.abi6690}{Sci. Adv. \textbf{8}, eabi6690 (2022)}.


\end{thebibliography}
\end{document}